\documentclass[twocolumn]{aastex61}
\usepackage{amsmath,amssymb,mathrsfs}

\newcommand\aastex{AAS\TeX}

\received{\today}
\revised{\***, 2018}
\accepted{\***,2018}
\submitjournal{ApJ, 859, 151}

\shorttitle{\aastex\ Magnetic field in IRDC G035.39-00.33}
\shortauthors{Liu et al.}

\shorttitle{A holistic perspective on the dynamics of G035.39-00.33 } \shortauthors{Liu et al.}

\begin{document}

\title{A holistic perspective on the dynamics of G035.39-00.33: the interplay between gas and magnetic fields }

\AuthorCollaborationLimit=200

\correspondingauthor{Tie Liu}
\email{liutiepku@gmail.com}

\author{Tie Liu}
\affiliation{Korea Astronomy and Space Science Institute, 776 Daedeokdaero, Yuseong-gu, Daejeon 34055, Republic of Korea}
\affiliation{East Asian Observatory, 660 N. A'ohoku Place, Hilo, HI 96720, USA}

\author{Pak Shing Li}
\affiliation{Astronomy Department, University of California, Berkeley, CA 94720}

\author{Mika Juvela}
\affiliation{Department of Physics, P.O.Box 64, FI-00014, University of Helsinki, Finland}

\author{Kee-Tae Kim}
\affiliation{Korea Astronomy and Space Science Institute, 776 Daedeokdaero, Yuseong-gu, Daejeon 34055, Republic of Korea}

\author{Neal J. Evans II}
\affiliation{Department of Astronomy, The University of Texas at Austin, 2515 Speedway, Stop C1400, Austin, TX 78712-1205}
\affiliation{Korea Astronomy and Space Science Institute, 776 Daedeokdaero, Yuseong-gu, Daejeon 34055, Republic of Korea}

\author{James Di Francesco}
\affiliation{NRC Herzberg Astronomy and Astrophysics, 5071 West Saanich Rd, Victoria, BC V9E 2E7, Canada}
\affiliation{Department of Physics and Astronomy, University of Victoria, Victoria, BC V8P 5C2, Canada}

\author{Sheng-Yuan Liu}
\affiliation{Institute of Astronomy and Astrophysics, Academia Sinica. 11F of Astronomy-Mathematics Building, AS/NTU No.1, Sec. 4, Roosevelt Rd, Taipei 10617, Taiwan}

\author{Jinghua Yuan}
\affiliation{National Astronomical Observatories, Chinese Academy of Sciences, Beijing, 100012, China}

\author{Ken'ichi Tatematsu}
\affiliation{National Astronomical Observatory of Japan, National Institutes of Natural Sciences, 2-21-1 Osawa, Mitaka, Tokyo 181-8588, Japan}

\author{Qizhou Zhang}
\affiliation{Harvard-Smithsonian Center for Astrophysics, 60 Garden Street, Cambridge, MA 02138, USA}

\author{Derek Ward-Thompson}
\affiliation{Jeremiah Horrocks Institute for Mathematics, Physics \& Astronomy, University of Central Lancashire, Preston PR1 2HE, UK}

\author{Gary Fuller}
\affiliation{UK ALMA Regional Centre Node, Jodrell Bank Centre for Astrophysics, School of Physics and Astronomy, The University of Manchester, Oxford Road, Manchester M13 9PL, UK}

\author{Paul F. Goldsmith}
\affiliation{Jet Propulsion Laboratory, California Institute of Technology, 4800 Oak Grove Drive, Pasadena, CA 91109, USA}

\author{P. M. Koch}
\affiliation{Institute of Astronomy and Astrophysics, Academia Sinica. 11F of Astronomy-Mathematics Building, AS/NTU No.1, Sec. 4, Roosevelt Rd, Taipei 10617, Taiwan}

\author{Patricio Sanhueza}
\affiliation{National Astronomical Observatory of Japan, National Institutes of Natural Sciences, 2-21-1 Osawa, Mitaka, Tokyo 181-8588, Japan}

\author{I. Ristorcelli}
\affiliation{IRAP, Universit\'{e} de Toulouse, CNRS, UPS, CNES, Toulouse, France}

\author{Sung-ju Kang}
\affiliation{Korea Astronomy and Space Science Institute, 776 Daedeokdaero, Yuseong-gu, Daejeon 34055, Republic of Korea}

\author{Huei-Ru Chen}
\affiliation{Institute of Astronomy and Department of Physics, National Tsing Hua University, Hsinchu, Taiwan}

\author{N. Hirano}
\affiliation{Institute of Astronomy and Astrophysics, Academia Sinica. 11F of Astronomy-Mathematics Building, AS/NTU No.1, Sec. 4, Roosevelt Rd, Taipei 10617, Taiwan}

\author{Yuefang Wu}
\affiliation{Department of Astronomy, Peking University, 100871, Beijing China}

\author{Vlas Sokolov}
\affiliation{Max Planck Institute for Extraterrestrial Physics, Gie{\ss}enbachstra{\ss}se 1, 85748, Garching bei M\"{u}nchen, Germany}

\author{Chang Won Lee}
\affiliation{Korea Astronomy and Space Science Institute, 776 Daedeokdaero, Yuseong-gu, Daejeon 34055, Republic of Korea}
\affiliation{Korea University of Science and Technology, 217 Gajeong-ro, Yuseong-gu, Daejeon 34113, Republic of Korea}

\author{Glenn J. White}
\affiliation{Department of Physics and Astronomy, The Open University, Walton Hall, Milton Keynes, MK7 6AA, UK}
\affiliation{RAL Space, STFC Rutherford Appleton Laboratory, Chilton, Didcot, Oxfordshire, OX11 0QX, UK}

\author{Ke Wang}
\affiliation{European Southern Observatory, Karl-Schwarzschild-Str.2, D-85748 Garching bei M\"{u}nchen, Germany}

\author{David Eden}
\affiliation{Astrophysics Research Institute, Liverpool John Moores University, IC2, Liverpool Science Park, 146 Brownlow Hill, Liverpool L3 5RF, UK}

\author{Di Li}
\affiliation{National Astronomical Observatories, Chinese Academy of Sciences, Beijing, 100012, China}
\affiliation{Key Laboratory of Radio Astronomy, Chinese Academy of Science, Nanjing 210008, China}

\author{Mark Thompson}
\affiliation{Centre for Astrophysics Research, School of Physics Astronomy \& Mathematics, University of Hertfordshire, College Lane, Hatfield, AL10 9AB, UK}

\author{Kate M Pattle}
\affiliation{Institute of Astronomy and Department of Physics, National Tsing Hua University, Hsinchu, Taiwan}

\author{Archana Soam}
\affiliation{Korea Astronomy and Space Science Institute, 776 Daedeokdaero, Yuseong-gu, Daejeon 34055, Republic of Korea}

\author{Evert Nasedkin}
\affiliation{Department of Physics and Astronomy, University of Waterloo, Waterloo, Ontario, N2L 3G1, Canada}

\author{Jongsoo Kim}
\affiliation{Korea Astronomy and Space Science Institute, 776 Daedeokdaero, Yuseong-gu, Daejeon 34055, Republic of Korea}

\author{Gwanjeong Kim}
\affiliation{National Astronomical Observatory of Japan, National Institutes of Natural Sciences, 2-21-1 Osawa, Mitaka, Tokyo 181-8588, Japan}

\author{Shih-Ping Lai}
\affiliation{Institute of Astronomy and Department of Physics, National Tsing Hua University, Hsinchu, Taiwan}

\author{Geumsook Park}
\affiliation{Korea Astronomy and Space Science Institute, 776 Daedeokdaero, Yuseong-gu, Daejeon 34055, Republic of Korea}

\author{Keping Qiu}
\affiliation{School of Astronomy and Space Science, Nanjing University, Nanjing 210023}

\author{Chuan-Peng Zhang}
\affiliation{National Astronomical Observatories, Chinese Academy of Sciences, Beijing, 100012, China}

\author{Dana Alina}
\affiliation{Department of Physics, School of Science and Technology, Nazarbayev University, Astana 010000, Kazakhstan}

\author{Chakali Eswaraiah}
\affiliation{Institute of Astronomy and Department of Physics, National Tsing Hua University, Hsinchu, Taiwan}

\author{Edith Falgarone}
\affiliation{LERMA, Observatoire de Paris, PSL Research University, CNRS, Sorbonne Universit\'es, UPMC Univ. Paris 06, Ecole normale sup\'erieure, 75005 Paris, France}

\author{Michel Fich}
\affiliation{Department of Physics and Astronomy, University of Waterloo, Waterloo, Ontario, N2L 3G1, Canada}

\author{Jane Greaves}
\affiliation{School of Physics and Astronomy, Cardiff University, Cardiff CF24 3AA, UK}

\author{Q.-L. Gu}
\affiliation{Department of Physics, The Chinese University of Hong Kong, Shatin, New Territory, Hong Kong, China}

\author{Woojin Kwon}
\affiliation{Korea Astronomy and Space Science Institute, 776 Daedeokdaero, Yuseong-gu, Daejeon 34055, Republic of Korea}
\affiliation{Korea University of Science and Technology, 217 Gajeong-ro, Yuseong-gu, Daejeon 34113, Republic of Korea}

\author{Hua-bai Li}
\affiliation{Department of Physics, The Chinese University of Hong Kong, Shatin, New Territory, Hong Kong, China}

\author{Johanna Malinen}
\affiliation{Institute of Physics I, University of Cologne, Z�lpicher Str. 77, D-50937, Cologne, Germany}

\author{Ludovic Montier}
\affiliation{IRAP, Universit\'{e} de Toulouse, CNRS, UPS, CNES, Toulouse, France}

\author{Harriet Parsons}
\affiliation{East Asian Observatory, 660 N. A'ohoku Place, Hilo, HI 96720, USA}

\author{Sheng-Li Qin}
\affiliation{Department of Astronomy, Yunnan University, and Key Laboratory of Astroparticle Physics of Yunnan Province, Kunming, 650091, China}

\author{Mark G. Rawlings}
\affiliation{East Asian Observatory, 660 N. A'ohoku Place, Hilo, HI 96720, USA}

\author{Zhi-Yuan Ren}
\affiliation{National Astronomical Observatories, Chinese Academy of Sciences, Beijing, 100012, China}

\author{Mengyao Tang}
\affil{Department of Astronomy, Yunnan University, and Key Laboratory of Astroparticle Physics of Yunnan Province, Kunming, 650091, China}

\author{Y.-W. Tang}
\affiliation{Institute of Astronomy and Astrophysics, Academia Sinica. 11F of Astronomy-Mathematics Building, AS/NTU No.1, Sec. 4, Roosevelt Rd, Taipei 10617, Taiwan}

\author{L. V. Toth}
\affiliation{E\"{o}tv\"{o}s Lor\'{a}nd University, Department of Astronomy, P\'{a}zm\'{a}ny P\'{e}ter s\'{e}t\'{a}ny 1/A, H-1117, Budapest, Hungary}

\author{Jiawei Wang}
\affiliation{Institute of Astronomy and Department of Physics, National Tsing Hua University, Hsinchu, Taiwan}

\author{Jan Wouterloot}
\affiliation{East Asian Observatory, 660 N. A'ohoku Place, Hilo, HI 96720, USA}

\author{H.-W. Yi}
\affiliation{School of Space Research, Kyung Hee University, Yongin-Si, Gyeonggi-Do 17104, Korea}

\author{H.-W. Zhang}
\affiliation{Department of Astronomy, Peking University, 100871, Beijing China}

\begin{abstract}
Magnetic field is one of the key agents that play a crucial role in shaping molecular clouds and regulating star formation, yet the complete information on the magnetic field is not well constrained due to the limitations in observations. We study the magnetic field in the massive infrared dark cloud G035.39-00.33 from dust continuum polarization observations at 850 $\micron$ with SCUBA-2/POL-2 at JCMT for the first time. The magnetic field tends to be perpendicular to the densest part of the main filament (F$_{M}$), whereas it has a less defined relative orientation in the rest of the structure, where it tends to be parallel to some diffuse regions. A mean plane-of-the-sky magnetic field strength of $\sim$50 $\mu$G for F$_{M}$ is obtained using Davis-Chandrasekhar-Fermi method. Based on $^{13}$CO (1-0) line observations, we suggest a formation scenario of F$_{M}$ due to large-scale ($\sim$10 pc) cloud-cloud collision. Using additional NH$_3$ line data, we estimate that F$_{M}$ will be gravitationally unstable if it is only supported by thermal pressure and turbulence. The northern part of F$_{M}$, however, can be stabilized by a modest additional support from the local magnetic field. The middle and southern parts of F$_{M}$ are likely unstable even if the magnetic field support is taken into account. We claim that the clumps in F$_{M}$ may be supported by turbulence and magnetic fields against gravitational collapse. Finally, we identified for the first time a massive ($\sim$200 M$_{\sun}$), collapsing starless clump candidate, ``c8", in G035.39-00.33. The magnetic field surrounding ``c8" is likely pinched, hinting at an accretion flow along the filament.
\end{abstract}


\section{Introduction}

The densest parts of massive molecular dark clouds are filamentary in form, with lengths ranging from several parsecs to more than 10 parsecs and with a width of a few tenths of a parsec \citep{and14,wang16}. One of the most striking results from Herschel observations in the Gould Belt clouds is the finding of an apparent characteristic width ($\sim$0.1 pc) of filamentary substructures \citep{and14}. The origin of such a characteristic width is not well understood. Projection effects or artifacts in the data analysis may also affect this result \citep{pan17}. However, numerical simulation modelling the interplay between turbulence, strong magnetic field, and gravitationally driven ambipolar diffusion are indeed able to reproduce filamentary structures with widths peaked at 0.1 pc over several orders of magnitude in column density \citep[e.g.,][]{Auddy16,fed16}. Therefore, it is crucial to investigate the interplay between turbulence, magnetic field, and gravity in filamentary clouds to understand their properties.

Statistical analysis of observed magnetic fields in the nearby Taurus, Musca, Ophiuchus, Chameleon, and Vela C molecular clouds as well as many infrared dark clouds (IRDCs) have revealed that the local magnetic fields tend to be perpendicular to the densest filaments, whereas the fields tend to be parallel in the lower density peripheries of those filaments \citep{chap11,cox16,planck16a,planck16b,sant16,mali16,alina17,sole17a,ward17,tang18a}.

Recent state-of-the-art large scale ideal magneto-hydrodynamic (MHD) simulations of the formation and structure of filamentary dark clouds suggest a complicated evolutionary process involving the interaction and fragmentation of dense, velocity coherent fibers into chains of cores \citep[e.g., ][]{klass17,lips17}. In the simulation of \cite{lips17}, the global magnetic field is roughly perpendicular to the long axis of the main filamentary cloud. Velocity coherent fibers are identified inside the filamentary cloud and appear to be intertwined along the main filamentary cloud. These results are similar to the structures identified in L1495/B213 \citep[see][]{hacar13,hacar16}. In 3-dimensional MHD simulations of cluster-forming turbulent molecular cloud clumps, \cite{klass17} find that magnetic fields are oriented more parallel to the major axis of the sub-virial clouds and more perpendicular in the denser and marginally bound clouds. Observationally, similar results are found by \cite{koch14} where the local angle ($|\delta|$) between an intensity gradient and a magnetic field orientation shows a possible bimodal distribution and clearly separates sub-critical from super-critical cores, based on 50 sources observed with the Submillimeter Array (SMA) and the Caltech Submillimeter Observatory (CSO).

Both numerical simulations \citep{lips15,lips17,klass17,sole17} and polarization observations \citep{gira13,chap11,li09,lihb15,koch12,koch14,qiu13,qiu14,zhang14,pillai15,cox16,ward17,pattle17} have found that the interstellar magnetic field is dynamically important to the formation of dense cores in filamentary clouds. It is, however, still unclear how important the magnetic field is in the formation of dense cores in filaments relative to the turbulence and gravity.

Optical or near-infrared absorption polarimetry that can trace the plane-of-the-sky (POS) projections of magnetic-field orientations has been limited to low-density, diffuse cloud material. Polarized sub-millimeter thermal dust emission, however, can trace magnetic fields in dense regions of clouds. Planck sub-millimeter polarimetry, while extensive, is limited to the study of distant clouds (e.g., IRDCs) due to the low angular resolution ($\sim5\arcmin$ or $\sim$1.5 pc at 1 kpc distance) \citep{planck16a,planck16b,alina17}. High angular resolution observations of polarized sub-millimeter thermal dust emission toward filamentary clouds are much better tracing the cores but are still very rare. Such observations, specifically of quiescent filamentary clouds that are not greatly affected by the star forming activities, are needed to explore the roles of magnetic field in dense core formation in filamentary clouds. One example of a massive but quiescent filamentary cloud is IRDC G035.39-00.33 (hereafter denoted as G035.39).

Located at a distance of 2.9 kpc \citep{simon06}, G035.39 is an IRDC with a total mass of $\sim$16700 M$_{\sun}$ \citep{kai13}. G035.39 contains massive, dense clumps as revealed by dense molecular line observations \citep{henshaw13,henshaw14,henshaw17,jim14,zhang17}. Kinematically identified substructures and resolved narrow (0.028$\pm$0.005 pc) fibers have been identified in G035.39 \citep{henshaw17}, indicating the existence of interacting velocity-coherent fibers similar to those discovered in L1495/B213. High CO depletion factors \citep[f$_{D}\sim5-10$;][]{jim14} and a high deuterium fractionation (D$_{N_{2}H^{+}}$) of N$_{2}$H$^{+}$ \citep[mean D$_{N_{2}H^{+}}=0.04\pm0.01$;][]{bar16} in the dense cores of G035.39 indicate that G035.39 is chemically evolved but has been relatively unaffected by the ongoing star forming activities. Indeed, the dense cores in this filament are either starless or are associated with very low luminosity ``Class 0"-like IR-quiet protostars \citep{Nguyen11}.

G035.39 is also known as a Planck Galactic Cold Clump (PGCC), PGCC G35.49-0.31 \citep{planck16c}. PGCCs are ideal targets for investigating the initial conditions of star formation and for studying the properties of filamentary clouds \citep{planck11a,planck11b,planck16c,juvela10,juvela12,mont15,rivera16,rivera17,wu12,liu12,liu13c,liu15,tat17,meng13,zhang16,yuan16}. G035.39 has been observed as part of the JCMT legacy survey program ``SCUBA-2 Continuum Observations of Pre-protostellar Evolution (SCOPE)", which targets $\sim$1000 PGCCs in 850 $\micron$ continuum and suitable for the investigation of the initial conditions of star formation in widely different Galactic environments \citep{kim17,tang18b,liu16c,juvela18a,liu18,Yi18,zhang18}. The ``SCOPE" survey has provided us thousands of dense clumps \citep[Eden et al. 2018, in preparation;][]{liu18} for these studies.

The magnetic field surrounding G035.39 may not be affected by star-forming activities (like outflows) and, therefore, G035.39 is an ideal target for polarization observations of initial conditions for the formation of IRDCs. To this end, we conducted a number of linear polarization observations of the dust continuum emission at 850 $\micron$ with the new POL-2 polarimeter, operating in
conjunction with the SCUBA-2 (Submillimetre Common-User Bolometer Array 2), at the James Clerk Maxwell Telescope (JCMT). The SCUBA-2/POL-2 observations of G035.39 serve as a pilot study of magnetic fields in ``SCOPE" objects. The kinematics of the structures in G035.39 is also investigated thoroughly from molecular line observations.

Our paper is organized as follows. In Section 2, we discuss our observations using SCUBA-2/POL-2 850 $\micron$ polarization continuum, together with other continuum data used to study the spectral energy distribution of G035.39. We also present our molecular line observations. In Section 3, we present the results from these observational data, and in Section 4 we discuss the implication of the data relevant to filamentary cloud formation induced by the cloud-cloud collision (Section 4.1), the origin of magnetic field geometry (Section 4.2), the gravitational stability of the filaments (Section 4.3), and the physical properties of clumps inside G035.39 (Section 4.4). We summarize our findings in Section 5.

\section{Observations}

\subsection{Polarized 850 $\micron$ continuum data}

The POS magnetic field is traced by polarized 850 $\micron$ continuum data obtained with the SCUBA-2/POL-2 instrument at the James Clerk Maxwell Telescope (JCMT). The SCUBA-2/POL-2 observations of G035.39 (project code: M17BP050; PI: Tie Liu) were conducted from 2017 June to 2017 November using a
version of the SCUBA-2 DAISY mapping mode \citep{holland13} optimized for POL-2 observations (POL-2 DAISY mapping mode)\citep{frib16}\footnote{http://www.eaobservatory.org/jcmt/instrumentation/continuum/scuba-2/pol-2/}. In total, 70 scans were conducted. The beam size of the JCMT at 850 $\micron$ is 14$\arcsec$.1.
The POL-2 DAISY scan pattern uses a scan speed of 8$\arcsec$/s (compared to 155$\arcsec$/s for a SCUBA-2 DAISY scan pattern) and a fully sampled circular region with a diameter of 12$\arcmin$, with a waveplate rotation speed of 2 Hz \citep{ward17}. The full description of the SCUBA-2/POL-2 instrument and the POL-2 observational mode can be found in \cite{frib16} and \cite{ward17}. Since only the central 3$\arcmin$ diameter region has an approximately uniform coverage in the POL-2 DAISY observations, we obtained two adjacent maps to cover G035.39. The central pointings of the two maps are R.A.(J2000)=18:57:07, DEC(J2000)=+02:11:30 and R.A.(J2000)=18:57:10, DEC(J2000)=+02:08:00.

Data reduction is performed using a python script called \textit{pol2map} written within the STARLINK/SMURF package \citep{chap13}, which is specific for submillimetre data reduction (much of it specific to the JCMT). The default pixel size in SCUBA-2/POL-2 observations is 4$\arcsec$, but the final data is gridded to 8$\arcsec$ pixels in \textit{pol2map} to improve sensitivity. The Stokes Q, U and I data are all reduced with a filtering out scale of 200$\arcsec$. The output polarization percentage values are debiased using the mean of their Q and U variances to remove statistical biasing in regions of low signal-to-noise \citep{kwon18,soam18}. The details of data reduction with \textit{pol2map} can be found in \cite{kwon18}. The final co-added maps have an rms noise of $\sim$1.5 mJy/beam. The polarization angle $\theta$ is measured as $\theta=0.5$ arctan($U/Q$). The angle increases from north towards east, following the IAU convention. Throughout this paper, the polarization orientations obtained are rotated by 90$\arcdeg$ to show the magnetic field orientation projected on the plane of sky.

\subsection{Continuum data}

We use SCUBA-2 Stokes I 450 $\micron$ and 850 $\micron$ continuum data obtained from the legacy survey program ``SCOPE" \citep{liu18} and Herschel archival data from the Hi-GAL project \citep{Molinari2010} to construct the pixel by pixel SEDs of the G035.39 field.

The SCUBA-2 observations were conducted on 2016 April 13 under better weather conditions than SCUBA-2/POL-2 observations in 2017. Therefore, the 450 $\micron$ data were also obtained. The beam sizes at 450 $\micron$ and 850 $\micron$ are 7$\arcsec$.9 and 14$\arcsec$.1, respectively. The pixel sizes are 2$\arcsec$ and 4$\arcsec$ at 450 $\micron$ and 850 $\micron$, respectively. The rms levels at 450 $\micron$ and 850 $\micron$ are $\sim$60 mJy~beam$^{-1}$ and $\sim$10 mJy~beam$^{-1}$, respectively.

We use the level 2.5 Herschel/SPIRE (250-500\,$\mu$m) maps available in the Herschel
Science Archive\footnote{http://archives.esac.esa.int/hsa}, using extended source calibration. The
resolutions of the original maps at 250 $\micron$, 350 $\micron$ and 500 $\micron$ are approximately 18.3$\arcsec$, 24.9$\arcsec$, and 36.3$\arcsec$, respectively.

\subsection{Line observations}

Large-scale C$^{18}$O (1-0) and $^{13}$CO (1-0) are used to study the kinematics of the G035.39's natal molecular cloud. The C$^{18}$O (1-0) and $^{13}$CO (1-0) mapping data are obtained from the legacy survey program ``TRAO Observations of PGCCs (TOP)" \citep{liu18}. The observations were conducted on 2017 March 17. The map size is $30\arcmin\times30\arcmin$. The center of those maps is R.A.(J2000)=18:57:10, DEC(J2000)=+02:10:00.  The FWHM  beam size ($\theta_{B}$) is 47$\arcsec$. The main beam efficiency ($\eta_{B}$) is 51\%. The system temperature during observations is 243 K. The OTF data were smoothed to 0.33 km~s$^{-1}$ and the baseline removed with Gildas/CLASS. The rms level is 0.15 K in antenna temperature (T$_{A}^{*}$) at a spectral resolution of 0.33 km~s$^{-1}$.

Single-pointing observational data of the HCO$^{+}$ (1-0), H$^{13}$CO$^{+}$ (1-0) and H$_2$CO ($2_{1,2}-1_{1,1}$) lines are used to investigate the dynamical status of a starless clump in G035.39. The data taking with the Korean VLBI Network (KVN) 21-m telescope \citep{kim11} in Tamna station were obtained on 2017 November 26 in its single-dish mode. The rest frequencies of HCO$^{+}$ (1-0), H$^{13}$CO$^{+}$ (1-0) and H$_2$CO ($2_{1,2}-1_{1,1}$) lines are 89.18852 GHz, 86.754288 GHz, and 140.83952 GHz, respectively. The pointing position is R.A.(J2000)=18:57:11.38, DEC(J2000)=+02:07:27.9. The main beam sizes at 86 GHz and 140 GHz are 32$\arcsec$ and 23$\arcsec$, respectively. The main beam efficiencies at 86 GHz and 140 GHz are 44\% and 36\%, respectively. The data is reduced with Gildas/CLASS. The spectral resolution for both the HCO$^{+}$ (1-0) and H$^{13}$CO$^{+}$ (1-0) lines is $\sim$0.11 km~s$^{-1}$. The spectral resolution for H$_2$CO ($2_{1,2}-1_{1,1}$) is 0.07 km~s$^{-1}$. The on-source times for the HCO$^{+}$ (1-0), H$^{13}$CO$^{+}$ (1-0) and H$_2$CO ($2_{1,2}-1_{1,1}$) observations are 10 minutes, 15 minutes, and 25 minutes, respectively. The system temperatures during observations are 184 K, 188 K and 171 K, respectively. The achieved rms levels in antenna temperatures are $\sim$0.05 K, $\sim$0.04 K and $\sim$0.03 K, respectively.

We also use the NH$_{3}$ (1,1) line data from \cite{soko17}. The GBT beam at NH$_{3}$ (1,1) line frequency is 32$\arcsec$. The details of the NH$_{3}$ (1,1) observations can be found in \cite{soko17}.

\section{Results}

\subsection{Structure and magnetic field geometry in G035.39}

\begin{figure*}[tbh!]
\centering
\includegraphics[angle=0,scale=1.2]{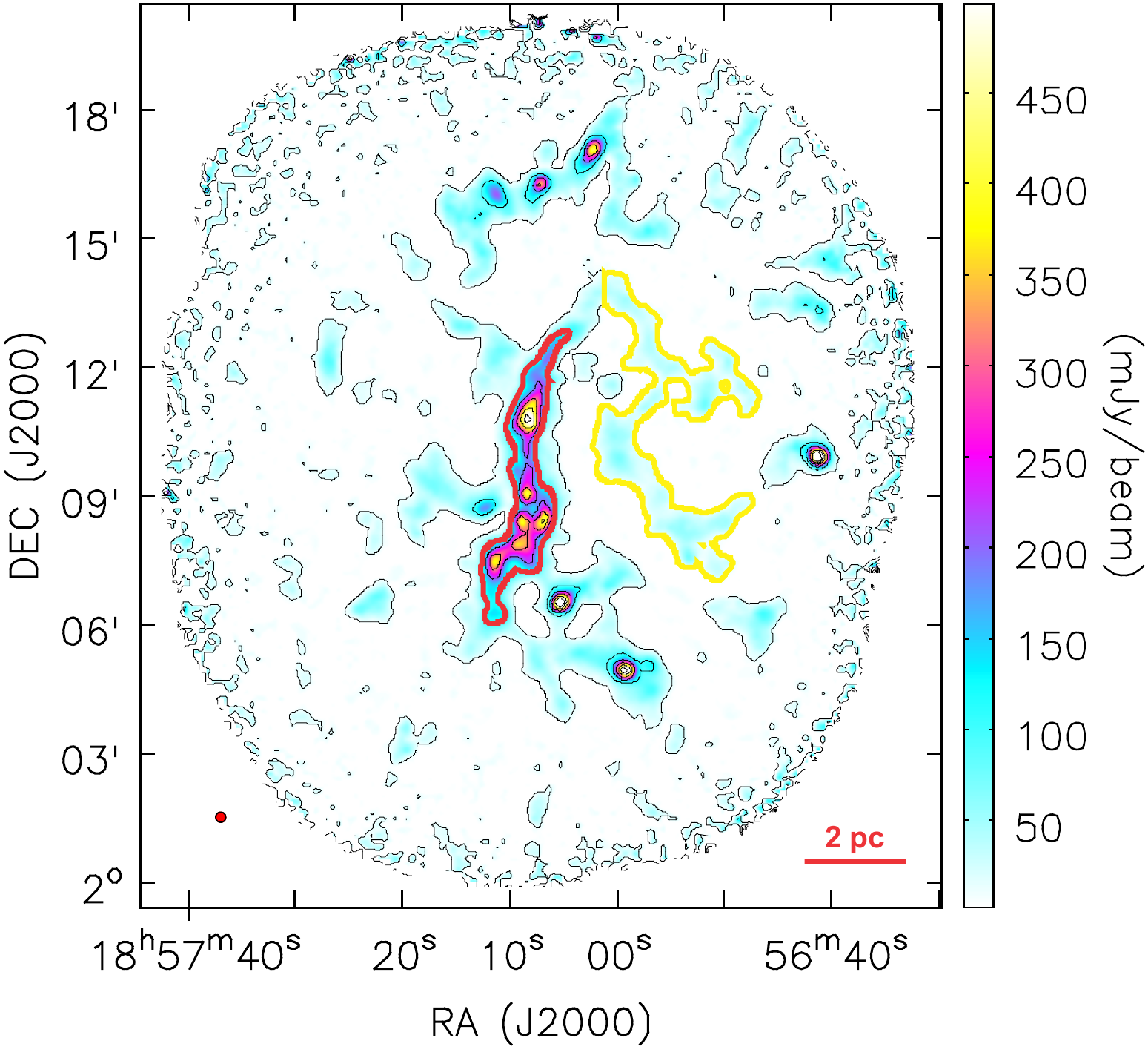}
\caption{The Stokes I image at 850 $\micron$ for G035.39. The outer contour level is 10 mJy~beam$^{-1}$. The inner contours are from 100 mJy~beam$^{-1}$ to 500 mJy~beam$^{-1}$ in steps of 100 mJy~beam$^{-1}$. The red contour (100 mJy/beam) outlines the main filament $F_{M}$ and the yellow contour (10 mJy/beam) outlines the faint western elongated structures $F_{W}$. The red filled circle corresponds to the beam size. \label{stokesI} }
\end{figure*}

\begin{figure*}[tbh!]
\centering
\includegraphics[angle=0,scale=0.9]{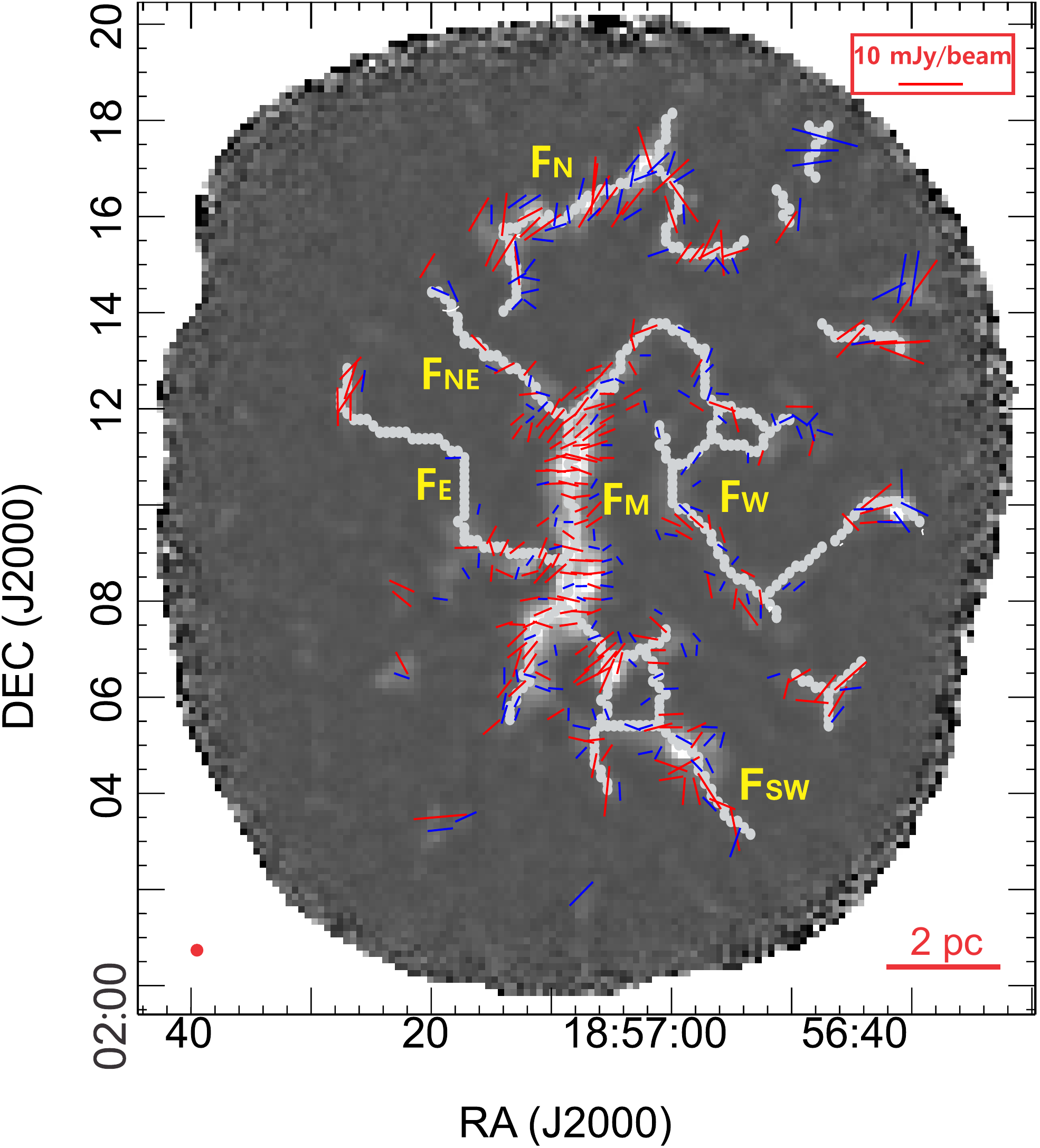}
\caption{JCMT/POL-2 map of G035.39. The background is the Stokes I image at 850 $\micron$. The magnetic field orientations are averaged within 16$\arcsec$ pixels. The red orientations are those detected at $SNR>$3 for polarization levels (P). The blue orientations are $2<SNR<3$ for P. The cutoff for Stokes I is $SNR>10$. The length of the orientations represents the polarization intensity in mJy~beam$^{-1}$ (see scale bar). The gray thick curves show the skeletons of the elongated structures. The red filled circle corresponds to the beam size. \label{cloud} }
\end{figure*}

Figure \ref{stokesI} shows the 850 $\micron$ Stokes I image. Besides the main filament (outlined with a red thick contour, hereafter denoted as \textit{$F_{M}$}) located at the center of the image, which was identified in previous work \citep{kai13}, the deep SCUBA-2/POL-2 observations reveal several fainter adjacent elongated structures (\textit{$F_{W}$}, \textit{$F_{SW}$}, \textit{$F_{E}$}, and \textit{$F_{NE}$}; see Figure \ref{cloud}) connected to \textit{$F_{M}$}. The skeletons of these elongated structures are identified by using the FILFINDER algorithm \citep{koch15} in 850 $\micron$ Stokes I emission above 3 $\sigma$ (1 $\sigma\sim$1.5 mJy~beam$^{-1}$). The skeletons are more easily identified in the high-contrast 850 $\micron$ Stokes I image than Herschel images because the extended diffuse emission is filtered out in SCUBA-2/POL-2 data. With larger filtering out scale, more extended emission can be recovered in 850 $\micron$ Stokes I data \citep{liu18}. Contamination from extended emission in 850 $\micron$ continuum will reduce the contrast between the skeletons and the background emission. The FILFINDER algorithm adopting the techniques of mathematical morphology not only can identify the bright filaments, but can also reliably extract a population of the faint filaments \citep{koch15}. The gray thick curves in Figure \ref{cloud} show the skeletons of the elongated structures.

\textit{$F_{M}$} has a length of $\sim$6.8 pc, measured from its skeleton. The longest elongated structure (outlined with a yellow thick contour in Figure \ref{stokesI}; denoted as \textit{$F_{W}$}) having a similar length ($\sim$6.7 pc) to \textit{$F_{M}$} is connected to the northern end of \textit{$F_{M}$}. The mean intensities of \textit{$F_{M}$} and \textit{$F_{W}$} at 850 $\micron$ within the 10 mJy~beam$^{-1}$ contours of the Stokes I image are $\sim$100 mJy~beam$^{-1}$ and $\sim$24 mJy~beam$^{-1}$, respectively, suggesting that \textit{$F_{W}$} is about four times fainter than \textit{$F_{M}$}.

The POS magnetic field orientations are shown in Figure \ref{cloud}. The field orientations are nearly perpendicular to the major axis of \textit{$F_{M}$} at the middle ridge, but tend to be parallel to its major axis at the lower density tails. The field orientations of the elongated structures (\textit{$F_{SW}$}, \textit{$F_{E}$}, and \textit{$F_{NE}$}) in their denser regions close to the junctions with \textit{$F_{M}$} also tend to be perpendicular to their skeletons. In contrast, the field orientations associated with \textit{$F_{W}$} are more parallel to its major axis. In this paper, we will mainly focus on \textit{$F_{M}$}. More detailed analysis and modeling of magnetic field geometry in the whole G035.39 field will be presented in a forthcoming paper (Juvela et al. 2018, in preparation).

\begin{figure*}[tbh!]
\centering
\includegraphics[angle=90,scale=0.6]{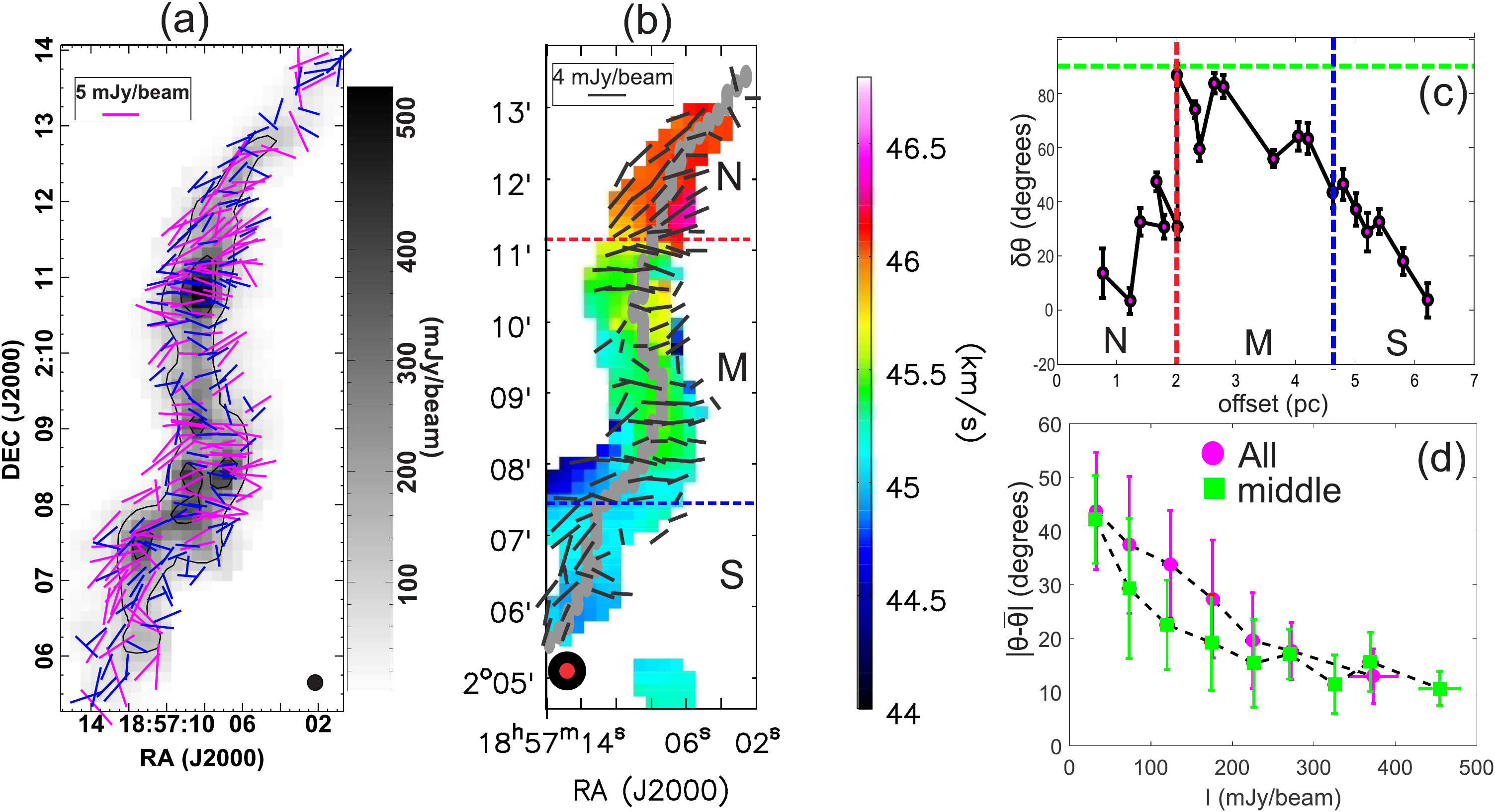}
\caption{(a). JCMT/POL-2 map of G035.39 main filament. The background is the Stokes I image at 850 $\micron$. The pixel size is the default value, i.e., 8$\arcsec$. The pink orientations are from polarization levels with $SNR>$3 . The blue orientations are from the polarization levels with $2<SNR<3$ . The cutoff for Stokes I is $SNR>10$. The length of the orientations represent the polarization intensity in mJy/beam. The contour levels are 100 mJy/beam for the outer contours, and 300 mJy/beam for the inner contours. The black filled circle corresponds to the beam size. (b). The magnetic field orientations of the G035.39 main filament are shown in black. The magnetic field orientations are averaged with a 16$\arcsec$ pixel boxcar filter. The background image is NH$_{3}$ centroid velocity map \citep{soko17}. The orientations are from polarization levels with $SNR>$2.  The cutoff for Stokes I is $SNR>10$. The three parts (N, M, S) of the main filament are divided by the blue and red dashed lines. The red and black filled circles correspond to the beam sizes of JCMT/POL-2 850 $\micron$ continuum and NH$_3$ (1,1), respectively. (c). The angle differences ($\delta\theta$) between an average field orientation (with a 32$\arcsec$ pixel boxcar filter) and the nearest skeleton change along the skeleton from the northern to the southern end. The three parts (N, M, S) of the main filament are divided by the blue and red vertical dashed lines. The horizontal green dashed line marks the $\delta\theta=90\arcdeg$. (d). The bin-averaged magnetic field position angle variance ($|\theta-\overline{\theta}|$) as a function of Stokes I intensity, subtracted by a mean position angle of ($\overline{\theta}=86\arcdeg$). The circles are bin-averaged angle variations of orientations along the main filament. The boxes are bin-averaged angle variations of orientations in the middle part of the main filament.  \label{main} }
\end{figure*}

Panel (a) in Figure \ref{main} shows the magnetic field orientations associated with only \textit{$F_{M}$}. The magnetic field orientations are more disordered at the two ends and near the edges of the filament. In contrast, the magnetic field orientations become more ordered along the central spine of the filament. We average the orientations with a 16$\arcsec$ pixel boxcar filter as \cite{pattle17} did and present the averaged orientations overlaid on a centroid velocity image of NH$_{3}$ (1,1) from \cite{soko17} in panel (b) of Figure \ref{main}. We divide \textit{$F_{M}$} into three regions (``N", ``M", ``S"), which show obvious differences in velocities and magnetic field geometries. ``N" and ``S" show redshifted and blueshiftied line-of-sight velocities with respect to ``M". In ``N" and ``S," the magnetic field orientations are more parallel to the filament skeletons. In contrast, the magnetic field orientations are more perpendicular to the filament skeletons in ``M."

\begin{figure}[tbh!]
\centering
\includegraphics[angle=0,scale=0.6]{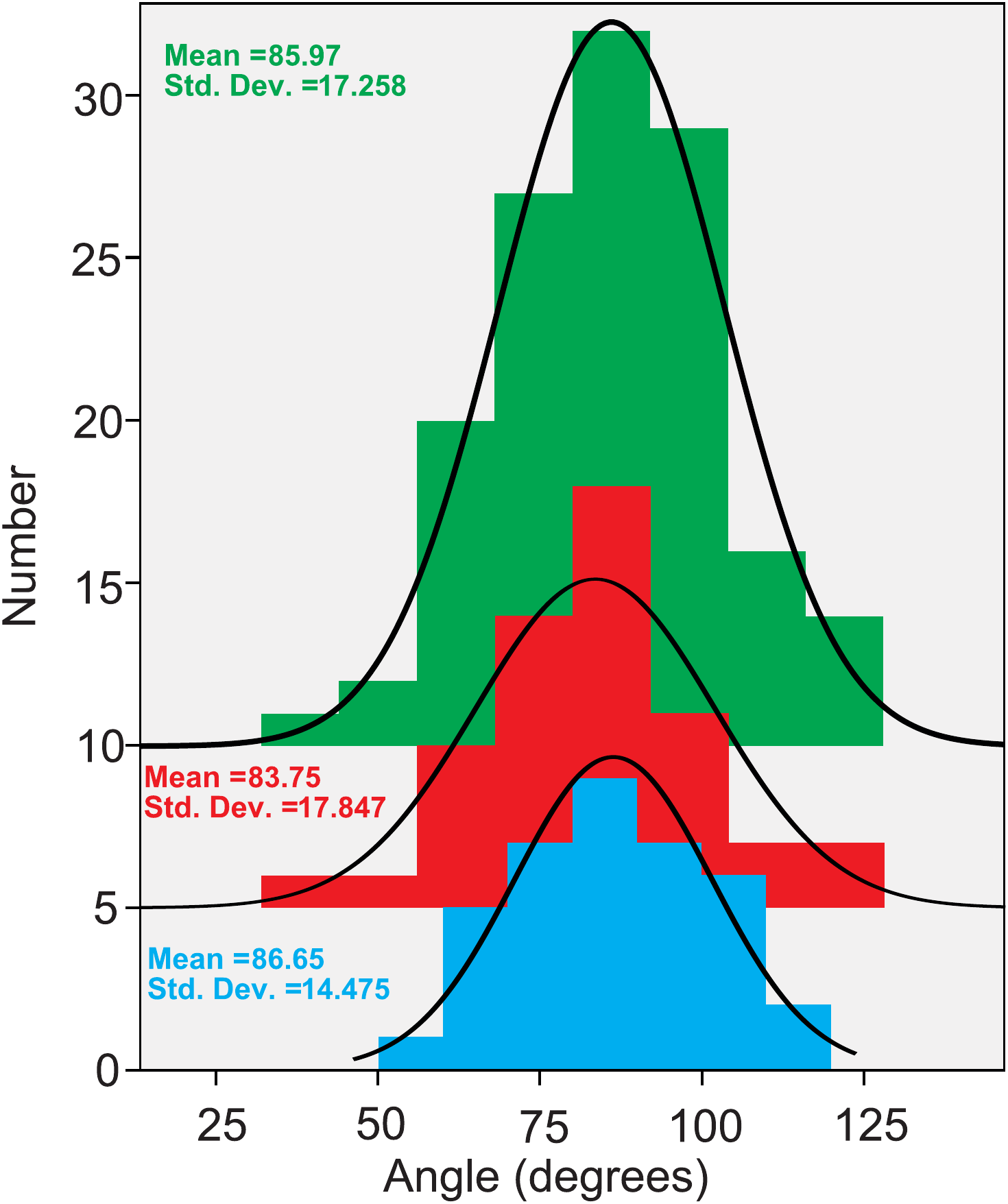}
\caption{Histograms of magnetic field orientations in the middle (``M") part of the main filament. Only orientations with Stokes I larger than 100 mJy/beam are included in statistics. The histogram of magnetic field orientations in the whole middle part region is shown in green color. The blue and red histograms are the magnetic field orientations in the sub-regions associated with dense clumps identified in section 3.3. The blue histogram is the position angles in the clump ``c3" region. The red histogram is the position angles in the region covering clumps ``c5, c6 and c7". The black lines are Gaussian fits. The means and standard deviations (std.dev.) in the plots are obtained from Gaussian fitting.  \label{stats} }
\end{figure}

To investigate how the ordered magnetic field orientations change along the filament, we average the magnetic field orientations with a 32$\arcsec$ pixel boxcar filter and calculate the angles ($\delta\theta$) between the mean magnetic field orientations and the local orientations of the nearest skeletons. The boxcar filter has a size of 32$\arcsec$ (or 0.45 pc), similar to the radius of the filament. The local orientations of the skeletons were measured from their gradients, similar to what \cite{koch12} did. The $\delta\theta$ as a function of offset distance from the northern end of \textit{$F_{M}$} is presented in panel (c) of Figure \ref{main}. The magnetic field orientations in the end regions (``N" and ``S") are more parallel to the major axis of the filament with $\delta\theta\leq40\arcdeg$. In contrast, the magnetic field orientations in the central part (``M") are more perpendicular to the major axis of the filament with $\delta\theta\geq60\arcdeg$. From each end, $\delta\theta$ increases toward the middle part of \textit{$F_{M}$}.

Figure \ref{stats} presents histograms of the magnetic field orientations in the whole ``M" region (green) and in the sub-regions (red and blue). The sub-regions are associated with dense clumps identified in section 3.3. The blue histogram shows the position angles in the clump ``c3" region. The red histogram shows the position angles in the region covering the clumps ``c5, c6 and c7". Only the orientations with Stokes I intensity larger than 100 mJy~beam$^{-1}$ are included. From a Gaussian fit, we obtain a mean magnetic field orientation of $\sim86\arcdeg$ and an orientation dispersion ($\sigma_{\theta}$) of $\sim17\arcdeg$ in the whole ``M" region. The mean magnetic field orientations ($\overline{\theta}$) in different sub-regions of the ``M" region are nearly the same, mostly perpendicular to the major axis of the filament.

Panel (d) of Figure \ref{main} shows how the magnetic field orientation variation (with a default pixel size of 8$\arcsec$) changes with Stokes I intensity. The magnetic field orientations ($\theta$) are subtracted by a mean value ($\overline{\theta}=86 \arcdeg$) in the densest part of ``M". Although the uncertainties of orientations caused by noise have been considered in this plot, the orientation variations in low density regions still show large dispersion, which can be improved by future higher sensitivity observations. The orientation variation $|\theta-\overline{\theta}|$, however, is nearly constant toward higher Stokes I intensity ($>$200 mJy~beam$^{-1}$), suggesting that the magnetic field orientations in the central part of the \textit{$F_{M}$} are more perpendicular to the major axis. In contrast, the orientation variations $|\theta-\overline{\theta}|$ becomes larger toward less dense regions (with Stokes I intensity smaller than 100 mJy~beam$^{-1}$), suggesting that magnetic field orientations in less central regions tend to be more parallel to the major axis.

\subsection{Properties of the main filament}

\begin{figure*}[tbh!]
\centering
\includegraphics[angle=90,scale=0.7]{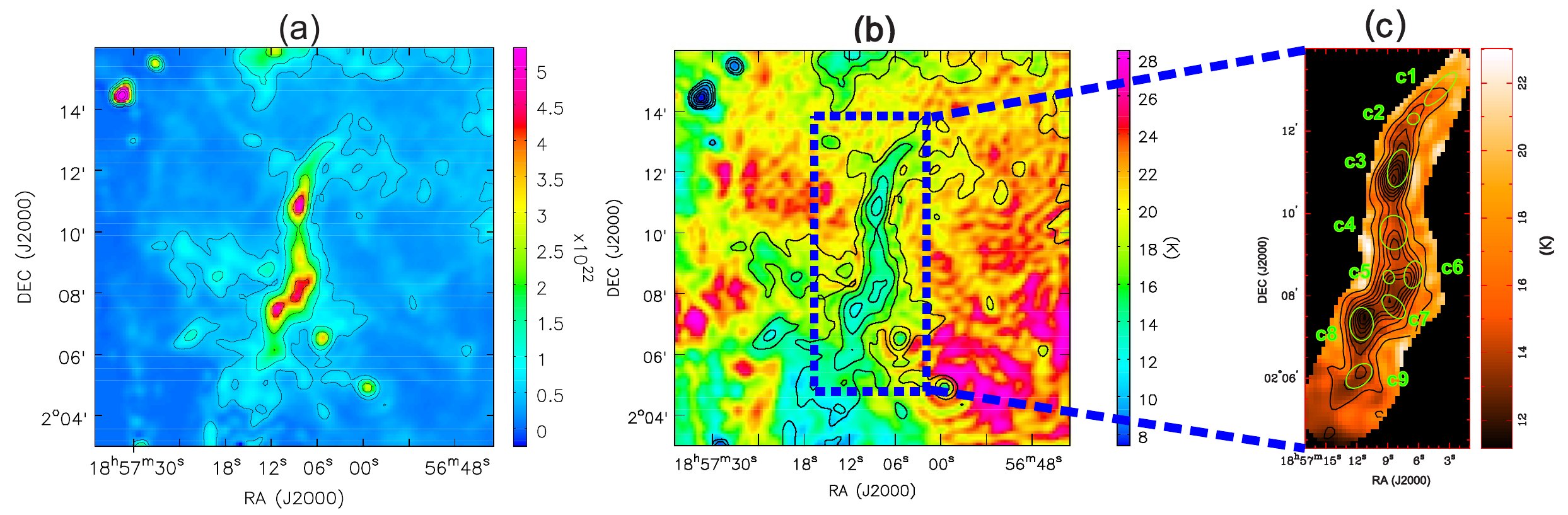}
\caption{(a) The H$_{2}$ column density map of the whole G035.39 field. The contours are [0.05, 0.1, 0.2, 0.4, 0.6, 0.8]$\times$1$\times10^{23}$ cm$^{-2}$. (b) The H$_{2}$ column density map is shown as contours overlaid on the dust temperature image. The contours are the same as in panel (a). (c) The column density map of the main filament F$_{M}$ is shown as contours overlaid on the dust temperature image. The contour levels start from 2.66$\times10^{21}$ cm$^{-2}$ in steps of 5.31$\times10^{21}$ cm$^{-2}$, which is 10\% of the peak value. The clumps identified via FELLWALKER are shown in green ellipses. \label{clumps} }
\end{figure*}

Column density ($\textit{N}_{\textrm{H2}}$) and dust temperature ($T_{\textbf{d}}$) maps of G035.39 are constructed from fitting the Herschel/SPIRE (250 $\micron$, 350 $\micron$, and 500 $\micron$) and JCMT/SCUBA-2 (450 $\micron$ and 850 $\micron$) data with a modified blackbody (MBB) function, assuming a dust emissivity spectral index ($\beta$) of 1.8 and the dust mass absorption coefficient $\kappa=0.1(\nu/1~\textrm{THz})^\beta$ cm$^2$g$^{-1}$ \citep{juvela18a}. Only SCUBA-2 data at 450 $\micron$ when the signal is above 150 MJy/sr and at 850 $\micron$ when the signal is above 30 MJy/sr in the \textit{$F_{M}$} are used. In such compact, bright regions, SCUBA-2 data are not much affected by spatial filtering. Outside such regions, the SCUBA-2 data is corrected with an offset obtained after comparing an image convolved to the 40$\arcsec$ resolution with a SPIRE-based prediction image at the same resolution.

The observations are fit with a model that consists of surface
brightness maps at reference wavelengths and of a color temperature map, all with a pixel size of 6$\arcsec$. The model is fit to the
observations as a global optimization problem. This involves the
convolution of the modified blackbody predictions of the model to the
resolution of each of the observed surface brightness maps.
After optimization, the final model is convolved to a resolution of
15$\arcsec$, which is close to the resolution of the SCUBA-2 data. More
details of the procedure are presented in a forthcoming paper (Juvela et
al., 2018, submitted). However, the results are found to be very
close to those that would be obtained with the method described in
Appendix A of \citet{Palmeirim2013}, which similarly tries to maximize
the resolution of the resulting column density maps. In our case, the
dust temperature is constrained mainly by the SPIRE data with the shortest
wavelength (250\,$\mu$m) being close to the peak of the spectrum of cold
dust emission. However, \cite{juvela12} estimated that even with 7\%
surface brightness errors, the SPIRE data gives a better than 1\,K
accuracy for the temperatures (for $T\sim$15\,K), which corresponds only to
$\sim$20\% uncertainty in the column density. In the fits, we used the 4\% and
10\% error estimates for SPIRE and SCUBA-2 data, respectively.

The $\textit{N}_{\textrm{H2}}$ and $T_{\textbf{d}}$ maps of G035.39 are presented in Figure \ref{clumps}. Although the background emission is subtracted, the above method is still subject to the usual caveats regarding to the line-of-sight temperature variations (Juvela et al. 2018, in preparation). The mean dust temperature of 15 K that we derived, however, is very consistent with the mean dust temperature of 14 K (from a small median filter method) and mean kinetic temperature of 13 K derived by \cite{soko17}.

The main filament \textit{$F_{M}$} has much higher column density and is colder than its surroundings. The $\textit{N}_{\textrm{H2}}$ and $T_{\textbf{d}}$ maps of \textit{$F_{M}$} are also presented in the right panel of Figure \ref{clumps}. Here we focus on the highest column density part of the main filament where $N_{H_2}>7\times10^{21}$ cm$^{-2}$, a column density threshold for core formation suggested in nearby clouds \citep{and14}. The mean column density of the sub-region is $\sim$1.8$\times10^{22}$ cm$^{-2}$. The length (\textit{L}) and the projected area (\textit{A}) of the ridge are $\sim$6.8 pc and $\sim$6.9 pc$^{2}$, respectively. Therefore, the total mass (\textit{M}) of the sub-region is calculated as:
\begin{equation}
M=N_{H_2}\times A\mu_{g}m_{H},
\end{equation}
where $\mu_{g}$=2.8 is the molecular weight per hydrogen molecule and m$_{H}$ is the mass of a hydrogen atom. The derived mass is $\sim$2800 M$_{\sun}$. Therefore, the line mass (mass per unit length) of the filament is M/L$\sim$410 M$_{\sun}$~pc$^{-1}$, which is within the range (223-635 M$_{\sun}$~pc$^{-1}$) given by \cite{soko17}. Assuming the filament has cylindrical geometry, the mean radius (\textit{r}) of the circular end of the cylinder is:
\begin{equation}
r=\frac{A}{2L}\approx0.5~\textrm{pc},
\end{equation}
where $A$ is the projected area of the filament. Therefore, the volume (\textit{V}) of the cylindrical filament is:
\begin{equation}
V=\pi r^2\times L\approx5.4~\textrm{pc}^{3},
\end{equation}
The mean volume density ($\textit{N}_{\textrm{H2}}$) is:
\begin{equation}
n_{H_2}=\frac{M}{V\mu_{g}m_{H}}\approx7.3\times10^3~\textrm{cm}^{-3}.
\end{equation}

In contrast, the average column density and dust temperature of the fainter, western elongated structures $F_{W}$ are $4.0\times10^{21}$ cm$^{-2}$ and $\sim$21 K, respectively, which are calculated within the 10 mJy~beam$^{-1}$ contours of the Stokes I intensity at 850 $\micron$. It is noted, however, that the column density and dust temperature of $F_{W}$ are less constrained than those of $F_{M}$ because only Herschel data are used in SEDs fit toward $F_{W}$. The mass of $F_{W}$ is $\sim$650 M$_{\sun}$. Considering the length ($\sim$6.7 pc) of $F_{W}$, its line mass and volume density are $\sim$100 M$_{\sun}$~pc$^{-1}$ and 1.7$\times10^3$ cm$^{-3}$, respectively. The total mass, line mass, column density and volume density of $F_{W}$ are all about 4 times smaller than those of $F_{M}$.

\subsection{Fragmentation of the main filament}

The main filament \textit{$F_{M}$} shows a chain of clumps with nearly even spacing. As shown in the right panel of Figure \ref{clumps}, we extracted nine dense clumps from the SCUBA-2 850 $\micron$ image
using the FELLWALKER \citep{berry15} source-extraction algorithm, which is a
part of the Starlink CUPID package \citep{berry07}. The core of the FELLWALKER algorithm is a gradient-tracing scheme that follow many different paths of steepest ascent in order to reach a significant
summit, each of which is associated with a clump \citep{berry07}. FELLWALKER is less dependent on specific parameter settings than other source-extraction algorithms \citep[e.g.,CLUMPFIND;][]{berry07}.

The source-extraction process with FELLWALKER is the same as that used by the JCMT Plane Survey and the details can be found in \cite{Moore15} and \cite{Eden17}. A mask constructed above a threshold of 3 $\sigma$ in the SNR map is applied to the intensity map as input for the task CUPID:EXTRACTCLUMPS, which extracts the peak and integrated flux-density values of the clumps. A further threshold for CUPID:FINDCLUMPS is the minimum number of contiguous pixels, which is set at 12 corresponding to the number of pixels expected to be found in an unresolved source with a peak SNR of 5 $\sigma$, given a 14$\arcsec$ beam and 4$\arcsec$ pixels.

The coordinates, radii (R$_{eff}$), mean dust temperature (T$_d$), mean column density ($N_{H_2}$), and mean non-thermal velocity dispersion ($\sigma_{NT}$) of th clumps derived from NH$_{3}$ (1,1) line emission are presented in Table \ref{clumppara}. The mean separation between clumps is $\sim$0.9 pc. The effective radii of clumps are $R_{eff}=\sqrt{ab}$, where \textit{a} and \textit{b} are the sizes of semi-major axis and semi-minor axis of the clump from 2D gaussian fit. The beam-deconvolved effective radii ($R_{eff}$) of clumps range from 0.12 pc to 0.31 pc with an average value of 0.23 pc. The dust temperatures of the clumps range from 13.2 K to 17.0 K with an average value of 14.6 K. Clump masses (M$_{clump}$) are derived as:
\begin{equation}
M_{clump}=\pi R_{eff}^2N_{H_2}\mu_g m_{H}
\end{equation}
Assuming that the clumps have spherical geometry, their H$_{2}$ volume density ($\textit{n}_{H_2}$) is derived as:
\begin{equation}
n_{H_2}=\frac{M}{\frac{4}{3}\pi R_{eff}^3\mu_g m_{H}}
\end{equation}
The volume densities and masses of clumps are also presented in Table \ref{clumppara}. The masses of clumps range from 16 M$_{\sun}$ to 219 M$_{\sun}$ with a mean value of $\sim$107 M$_{\sun}$. The volume density of clumps range from 8$\times10^3$ cm$^{-3}$ to 6.4$\times10^4$ cm$^{-3}$ with a mean value of 3$\times10^4$ cm$^{-3}$.

\begin{deluxetable*}{cccccccccccccc}[tbh!]
\centering
\tablecolumns{14} \tablewidth{0pc}\setlength{\tabcolsep}{0.05in}
\tablecaption{Parameters of dense clumps in the main filament \label{clumppara}}\tablehead{\colhead{Clump} & \colhead{RA} & \colhead{DEC}  & \colhead{R$_{eff}$\tablenotemark{a}} & \colhead{$T_{\textbf{d}}$\tablenotemark{b}} & \colhead{$N_{H_2}$\tablenotemark{c}}  & \colhead{$\sigma_{NT}$\tablenotemark{d}}  & \colhead{n$_{H_{2}}$} & \colhead{B$_{clump}$} & \colhead{$\sigma_{A}$} & \colhead{$\mathcal{M_A}$} & \colhead{M$_{clump}$}   &  \colhead{M$_{vir}$} &  \colhead{M$_{vir}^{B}$}\\
\colhead{No.}  & \colhead{(J2000)} & \colhead{(J2000)} & \colhead{(pc)}  & \colhead{(K)} & \colhead{(10$^{22}$ cm$^{-2}$)}  & \colhead{(km~s$^{-1}$)} & \colhead{(10$^{4}$ cm$^{-3}$)} & \colhead{($\mu$G)}  & \colhead{(km~s$^{-1}$)}& & \colhead{(M$_{\sun}$)} & \colhead{(M$_{\sun}$)} & \colhead{(M$_{\sun}$)} }
\startdata
c1  &  18:57:03.84  & +02:13:01.2  & 0.25 & 17.0 & 0.8 & 0.31  & 0.8 & 56  & 0.8 & 0.6 & 35  &  45  & 78   \\
c2  &  18:57:06.48  & +02:12:18.0  & 0.12 & 14.6 & 1.6 & 0.30  & 3.2 & 142 & 1.0 & 0.5 & 16  &  19  & 44     \\
c3  &  18:57:07.92  & +02:11:06.0  & 0.29 & 14.7 & 3.7 & 0.47  & 3.1 & 138 & 1.0 & 0.8 & 219 &  92  & 150     \\
c4  &  18:57:08.40  & +02:09:32.4  & 0.31 & 15.0 & 2.2 & 0.45  & 1.7 & 94  & 0.9 & 0.8 & 149 &  92  & 144     \\
c5  &  18:57:08.88  & +02:08:27.6  & 0.13 & 14.0 & 3.4 & 0.70  & 6.4 & 219 & 1.1 & 1.1 & 40  &  81  & 114     \\
c6  &  18:57:06.48  & +02:08:31.2  & 0.23 & 15.4 & 2.8 & 0.47  & 3.0 & 133 & 1.0 & 0.8 & 104 &  73  & 119      \\
c7  &  18:57:08.40  & +02:07:44.4  & 0.22 & 14.0 & 3.6 & 0.50  & 4.0 & 162 & 1.1 & 0.8 & 123 &  76  & 124     \\
c8  &  18:57:11.52  & +02:07:19.2  & 0.28 & 13.2 & 3.6 & 0.39  & 3.1 & 138 & 1.0 & 0.7 & 199 &  64  & 120  \\
c9  &  18:57:11.76  & +02:06:03.6  & 0.25 & 13.9 & 1.7 & 0.34  & 1.7 & 91  & 0.9 & 0.6 & 75  &  45  & 89   \\
\enddata
\tablenotetext{a}{$R_{eff}=\sqrt{ab}$, where a and b are the sizes of semi-major axis and semi-minor axis of the clump from 2D gaussian fit. $R_{eff}$ has been deconvolved with the beam.}
\tablenotetext{b}{clump-averaged dust temperature ($T_{\textbf{d}}$) }
\tablenotetext{c}{mean column density}
\tablenotetext{d}{clump-averaged velocity dispersion derived from linewidths of NH$_{3}$ (1,1) emission \citep{soko17}. }
\end{deluxetable*}

\subsection{Large-scale distribution of the gas}

\begin{deluxetable*}{ccccccccccc}[tbh!]
\centering
\tablecolumns{11} \tablewidth{0pc}\setlength{\tabcolsep}{0.05in}
\tablecaption{Parameters of molecular lines \label{linepara}}\tablehead{\colhead{Line} & \colhead{Velocity} & \colhead{FWHM} & \colhead{T$_{b}$} & \colhead{$T_{\textbf{d}}$} & \colhead{$\textit{N}_{\textrm{H2}}$}  & \colhead{n} & \colhead{$N_{\textrm{line}}$}  & \colhead{X$_{line}$} & \colhead{$\tau$} & \colhead{T$_{ex}$}\\
\colhead{}  & \colhead{(km~s$^{-1}$)} & \colhead{(km~s$^{-1}$)} & \colhead{(K)} & \colhead{(K)}  & \colhead{(cm$^{-2}$)} & \colhead{(cm$^{-3}$)}& \colhead{(cm$^{-2}$)}& \colhead{} & \colhead{} & \colhead{(K)}}
\startdata
\multicolumn{11}{c}{F$_{M}$ within 100 mJy~beam$^{-1}$ contour of 850 $\micron$ continuum}\\
\cline{1-11}
$^{13}$CO (1-0) & 44.9$\pm$0.1 & 3.3$\pm$0.1 & 4.9$\pm$0.1 & 14$\pm$1 & 2.4$\times10^{22}$ & 1.0$\times10^{4}$ & 2.2$\times10^{16}$ & 9.2$\times10^{-7}$ & 0.6 & 14.1\\
                & 13.1$\pm$0.1 & 2.2$\pm$0.2 & 1.3$\pm$0.1 \\
                & 27.2$\pm$0.1 & 1.7$\pm$0.1 & 2.0$\pm$0.1 \\
                & 56.2$\pm$0.1 & 4.3$\pm$0.2 & 1.6$\pm$0.1 \\
C$^{18}$O (1-0) & 44.9$\pm$0.1 & 2.1$\pm$0.1 & 1.2$\pm$0.1 & 14$\pm$1 & 2.4$\times10^{22}$ & 1.0$\times10^{4}$ & 2.6$\times10^{15}$ & 1.1$\times10^{-7}$ & 0.1 & 15.0\\
\cline{1-11}
\multicolumn{11}{c}{F$_{W}$ within 10 mJy~beam$^{-1}$ contour of 850 $\micron$ continuum}\\
\cline{1-11}
$^{13}$CO (1-0) & 45.4$\pm$0.1 & 3.9$\pm$0.3 & 2.4$\pm$0.1 & 21$\pm$1 & 4.0$\times10^{21}$ & 1.7$\times10^{3}$ & 8.5$\times10^{15}$ & 2.1$\times10^{-6}$ & 0.2 & 14.5\\
                & 13.0$\pm$0.1 & 2.5$\pm$0.3 & 1.3$\pm$0.1 \\
                & 27.9$\pm$0.2 & 2.6$\pm$0.5 & 0.9$\pm$0.1 \\
                & 93.1$\pm$0.2 & 2.3$\pm$0.4 & 1.1$\pm$0.1 \\
C$^{18}$O (1-0) & 44.9$\pm$0.1 & 1.8$\pm$0.4 & 0.3$\pm$0.1 & 21$\pm$1 & 4.0$\times10^{21}$ & 1.7$\times10^{3}$ & 4.5$\times10^{14}$ & 1.1$\times10^{-7}$ & 0.03 & 14.3\\
\enddata
\end{deluxetable*}

\begin{figure*}[tbh!]
\centering
\includegraphics[angle=-90,scale=0.65]{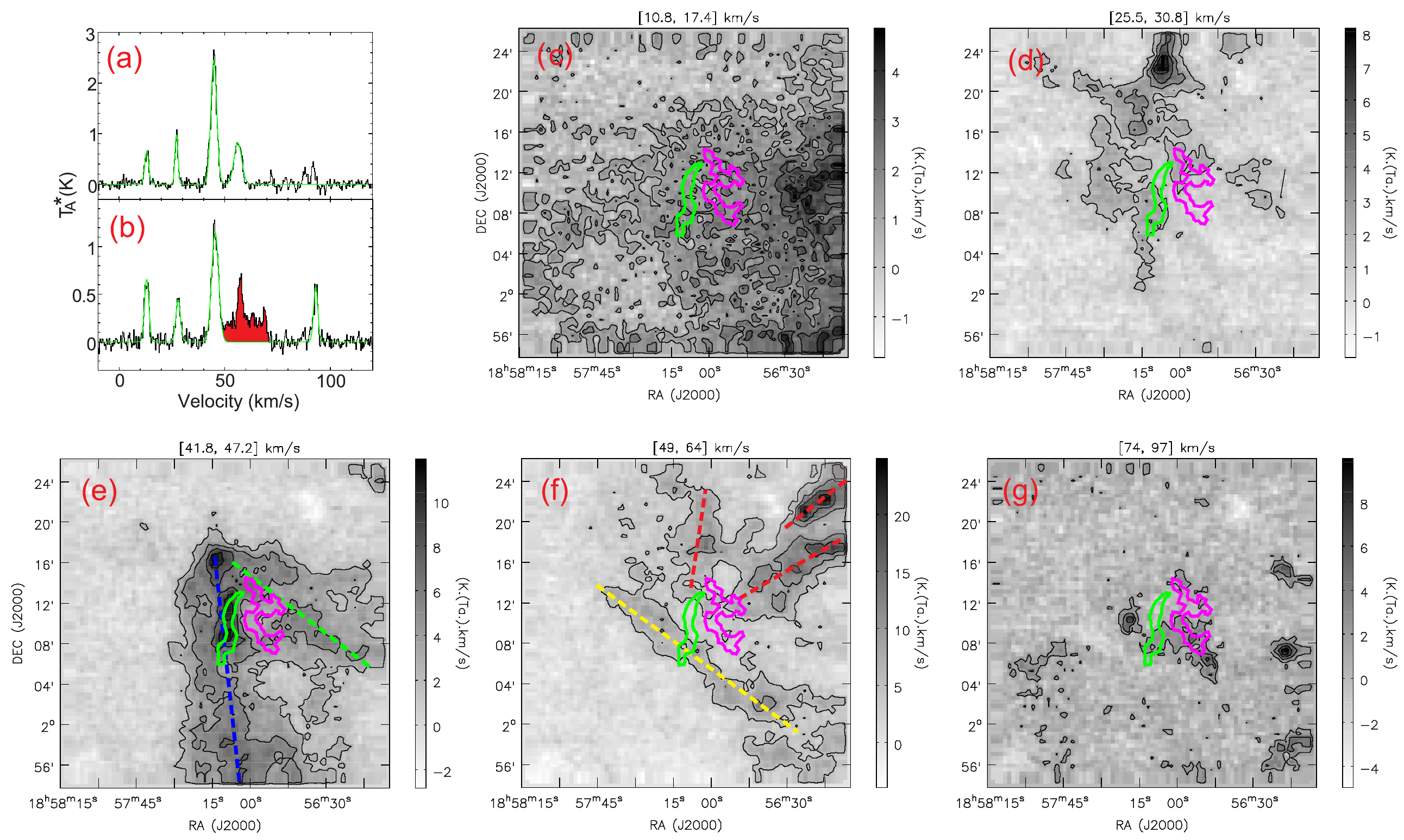}
\caption{Panel (a): Spectrum of $^{13}$CO (1-0) line emission averaged over the main filament ($F_{M}$) region. The green line shows the multi-component Gaussian fit. Panel (b): Spectrum of $^{13}$CO (1-0) line emission averaged over the western elongated structures ($F_{W}$) region. The green line shows the multi-component Gaussian fit. The red area shows the broad component which is not fitted.  Panels (c) to (g) show the integrated intensity of $^{13}$CO (1-0) line emission of different velocity components. The velocity intervals are shown above the image boxes. The contours are from 20\% to 80\% in steps of 20\% of peak values. The green and magenta thick contours mark the positions of the $F_{M}$ and $F_{W}$, respectively. The peak values in panels c, d, e, f and g are 4.9 K km~s$^{-1}$, 8.2 K km~s$^{-1}$, 11.9 K km~s$^{-1}$, 25.0 K km~s$^{-1}$, and 10.0 K km~s$^{-1}$, respectively. The dashed lines in panels (e) and (f) outline the elongated structures in the emission. \label{13CO} }
\end{figure*}

\begin{figure*}[tbh!]
\centering
\includegraphics[angle=-90,scale=0.65]{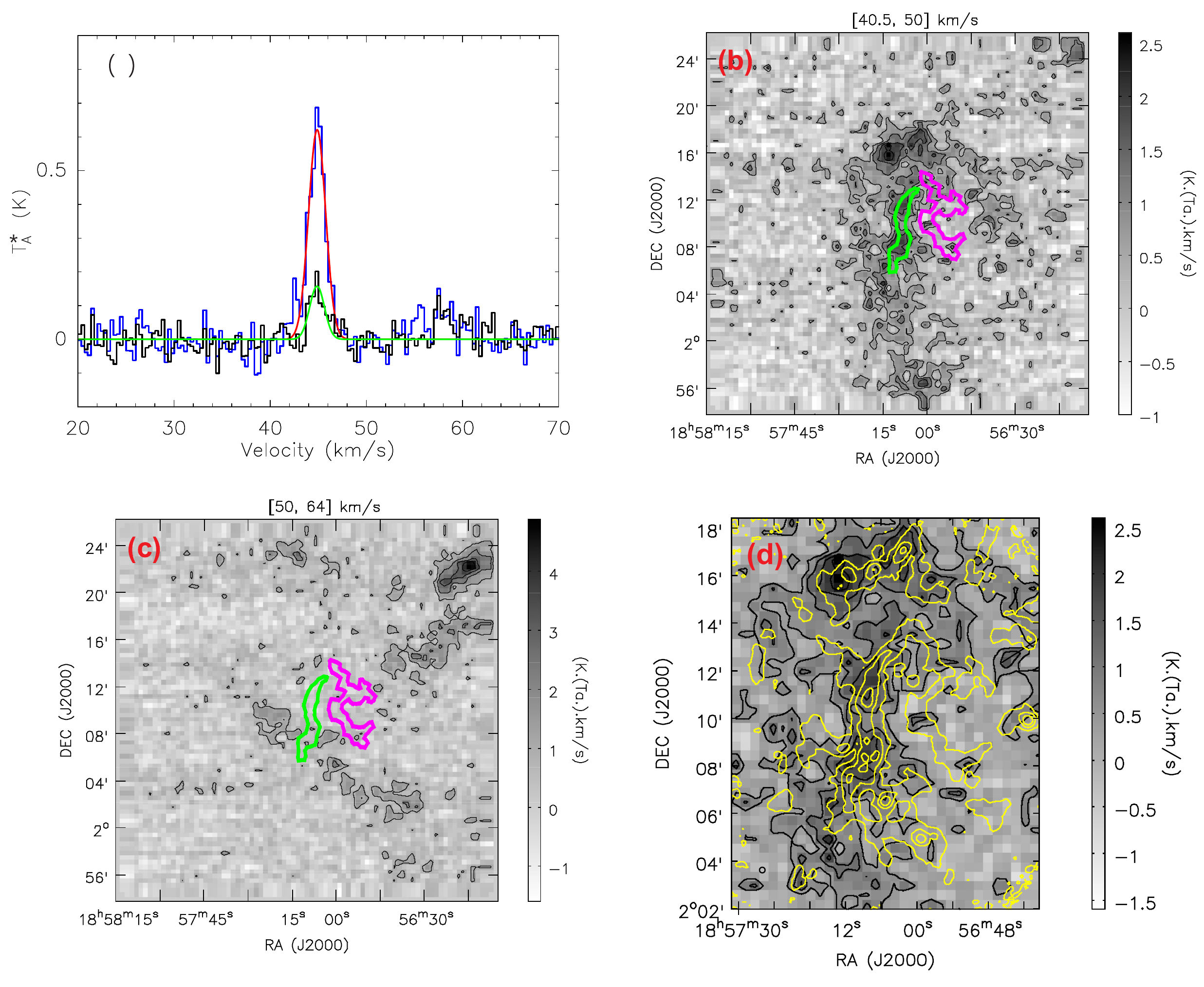}
\caption{(a): The spectra of C$^{18}$O (1-0) line emission averaged over $F_{M}$ and $F_{W}$ are shown in blue and black, respectively. The red and green curves show Gaussian fits. Panels (b) and (c) show the integrated intensity of C$^{18}$O (1-0) line emission of different velocity components. The velocity intervals are shown above the image boxes. The contours are from 20\% to 80\% in steps of 20\% of peak values. The peak values in panels (b) and (c) are 2.6 K km~s$^{-1}$  and 4.9 K km~s$^{-1}$, respectively. The green and magenta thick contours mark the positions of the $F_{M}$ and $F_{W}$, respectively.  (d): The gray image and black contours show the integrated intensity of C$^{18}$O (1-0) line emission as the same in panel (b). The yellow contours show 850 $\micron$ continuum emission. The contours are [0.15,0.5,1,2,3,4,5,6]$\times$ 100 mJy~beam$^{-1}$. \label{C18O} }
\end{figure*}

Panel (a) of Figure \ref{13CO} shows the averaged spectrum of the $^{13}$CO (1-0) line emission toward the main filament $F_{M}$ (enclosed by the green contour in panels (c) to (g) of Figure \ref{13CO}).  Multiple velocity components are seen in $^{13}$CO (1-0) line emission, suggesting that several clouds are overlapped along the line of sight. Those velocity components are well separated by at least 10 km~s$^{-1}$. The four strongest components can be well fitted with Gaussian functions. Their peak velocities, line widths and brightness temperature are listed in Table \ref{linepara}. The emission around 90 km~s$^{-1}$ is weak and was not fitted.

Panel (b) of Figure \ref{13CO} shows the averaged spectrum of the $^{13}$CO (1-0) line emission toward the western elongated structures $F_{W}$ (enclosed by the magenta contour in panels (c) to (g) of Figure \ref{13CO}). Five velocity components are seen in the $^{13}$CO (1-0) line emission. We fit four components with Gaussian functions and summarize the fitting parameters in Table \ref{linepara}. The emission (highlighted in red) between 50 km~s$^{-1}$ and 70 km~s$^{-1}$ contains a narrow component and a broad component. The broad component overlaps in velocity with the 45.4 km~s$^{-1}$ component.

Panels (c) to (g) of Figure \ref{13CO} present the integrated intensity maps of each velocity component in $^{13}$CO (1-0) line emission. The \textbf{structures} identified in 850 $\micron$ continuum emission are mainly associated with the velocity feature at 45 km~s$^{-1}$ (see panel e). Previous molecular line studies have also suggested that the main filament $F_{M}$ is dominated by emission around 45 km~s$^{-1}$ \citep{sanh12,jim14,henshaw13,henshaw14,henshaw17,soko17}. The emission around 13 km~s$^{-1}$ (see panel c) is very diffuse and is mainly distributed to the west of $F_{M}$. The emission around 27 km~s$^{-1}$ (see panel d) is mainly located to the north of $F_{M}$ and is partially overlapped with $F_{M}$. The emission around 93 km~s$^{-1}$ (see panel g) is overlapped with $F_{M}$ and $F_{W}$ but shows a very large velocity difference with respect to the 45 km~s$^{-1}$ component, suggesting that the 93 km~s$^{-1}$ component is not really physically associated with $F_{M}$ and $F_{W}$. Similar to the 45 km~s$^{-1}$ component, the emission around 56 km~s$^{-1}$ (see panel f) also shows filamentary structures. Its emission, however, is mainly distributed in the north and north-west of $F_{M}$ and $F_{W}$. A long filament (marked by yellow dashed line in panel f) in the 56 km~s$^{-1}$ cloud overlaps with the southern part of $F_{M}$. Since the velocity difference between the 56 km~s$^{-1}$ component and the 45 km~s$^{-1}$ component is larger than 10 km~s$^{-1}$, $F_{M}$ and $F_{W}$ may not be greatly affected by the 56 km~s$^{-1}$ cloud. The 56 km~s$^{-1}$ cloud, however, is adjacent to the west part (marked by the green dashed line in panel (e) of the 45 km~s$^{-1}$ cloud and shows broad emission therein (see panel b), which suggests that the 56 km~s$^{-1}$ cloud may be interacting with the 45 km~s$^{-1}$ cloud. The 45 km~s$^{-1}$ cloud shows two elongated structures as shown by the dashed lines in panel (e). The main part which contains $F_{M}$ has a length of at least $\sim$20 pc. The western part which contains $F_{W}$ has a length of $\sim$17 pc.

Panel (a) of Figure \ref{C18O} presents the averaged spectra of C$^{18}$O (1-0) line emission in the $F_{M}$ and $F_{W}$ regions. The C$^{18}$O (1-0) line emission also shows two velocity components at $\sim$45 km~s$^{-1}$ (see panel b) and $\sim$56 km~s$^{-1}$ (see panel c). The $\sim$56 km~s$^{-1}$ component, however, is very weak and marginally detected toward $F_{M}$ and $F_{W}$. Panels (b) and (c) present the integrated intensity maps of the two velocity components. $F_{M}$ is clearly associated with the 45 km~s$^{-1}$ emission but offset from the $\sim$56 km~s$^{-1}$ emission. As shown in panel (d), the C$^{18}$O (1-0) line emission around 45 km~s$^{-1}$ has similar morphology as the 850 $\micron$ continuum emission. \textit{$F_{W}$} is not obviously seen in C$^{18}$O (1-0) emission maps. The C$^{18}$O (1-0) emission signal is marginally detected as seen from the averaged spectrum in panel (a). From Gaussian fitting, the C$^{18}$O (1-0) line luminosity of \textit{$F_{W}$} is about 5 times smaller than that of \textit{$F_{M}$}. \textit{$F_{W}$} and \textit{$F_{M}$}, however, show very similar velocity ($\sim$44.9$\pm$0.1 km~s$^{-1}$ for both) and line widths (1.8$\pm$0.4 km~s$^{-1}$ for \textit{$F_{W}$} and 2.1$\pm$0.1 km~s$^{-1}$ for \textit{$F_{M}$} ) in C$^{18}$O (1-0) emission, suggesting that \textit{$F_{W}$} and \textit{$F_{M}$} are kinematically and spatially connected.

\textit{$F_{W}$} has $^{13}$CO (1-0) and C$^{18}$O (1-0) line widths similar to those of \textit{$F_{M}$}. \textit{$F_{W}$}, however, is about four times less dense than \textit{$F_{M}$}, suggesting that turbulence in \textit{$F_{W}$} likely plays a relatively more important role compared to \textit{$F_{M}$}.

Table \ref{linepara} summarizes the parameters of the averaged spectra of $^{13}$CO (1-0) and C$^{18}$O (1-0) line emission toward \textit{$F_{M}$} and \textit{$F_{W}$}, including the centroid velocities, line widths and peak brightness temperature from the Gaussian fitting. Their mean dust temperatures, column densities, and H$_{2}$ number densities are also listed. The column densities of $^{13}$CO (1-0) and C$^{18}$O (1-0) lines are derived using the non-LTE radiation transfer code, RADEX \citep{van07}. The inputs for RADEX are kinetic temperature ($T_{\textrm{k}}$), H$_{2}$ number density ($n_{\textrm{H}_\textrm{2}}$), molecular line column density ($N_{\textrm{line}}$), and line width ($\Delta v$). We assume that the $T_{\textrm{k}}$ is equal to the dust temperature. By fixing $T_{\textrm{k}}$, $n_{\textrm{H}_\textrm{2}}$, and $\Delta v$, we can calculate the brightness temperature for a grid of $N_{\textrm{line}}$ values and compare the model brightness temperature with observed values to find out the best $N_{\textrm{line}}$. The derived molecular line column densities and abundances are listed also in Table \ref{linepara}. The mean column densities of $^{13}$CO in \textit{$F_{M}$} and \textit{$F_{W}$} are $\sim2.2\times10^{16}$ cm$^{-2}$ and $\sim8.5\times10^{15}$ cm$^{-2}$, respectively. The mean column densities of C$^{18}$O in \textit{$F_{M}$} and \textit{$F_{W}$} are $\sim2.6\times10^{15}$ cm$^{-2}$ and $\sim4.5\times10^{14}$ cm$^{-2}$, respectively. The $^{13}$CO abundance in \textit{$F_{M}$} is about two times smaller than that in \textit{$F_{W}$}, possibly due to larger opacity in $^{13}$CO (1-0) emission in \textit{$F_{M}$}. In contrast, \textit{$F_{M}$} and \textit{$F_{W}$} have very similar C$^{18}$O abundances. By comparing the observed C$^{18}$O abundance ($X_{obs}$) with a canonical C$^{18}$O abundance ($X_{cano}\sim6.1\times10^{-7}$), \cite{jim14} claimed high CO depletion factors ($f_{D}=X_{cano}/X_{obs}\sim$5-12) in all three regions of \textit{$F_{M}$}. We derive a mean CO depletion factor of $\sim$6 for both $F_{M}$ and $F_{W}$ if we adopt the same canonical C$^{18}$O abundance.
$F_{M}$ is much denser than $F_{W}$ and thus CO may be expected to more depleted in $F_{M}$. The C$^{18}$O and $^{13}$CO abundances in both $F_{M}$ and $F_{W}$, however, only vary by a factor of 1-2 in our observations, suggesting that CO in $F_{M}$ may not be depleted as severely as suggested in previous studies \citep[e.g.,][]{jim14}. Observations in \cite{jim14} have much better spatial resolution than ours and thus they arguably trace better the inner parts of the filament. Therefore, high CO depletion may occur in densest regions of the filament. The uncertainties of the adopted canonical abundance, however, may prevent an accurate determination of CO depletion levels because observed CO abundances vary cloud-by-cloud in the Galaxy \citep{liu13c,Gian14}. Finally, the optical depths and excitation temperatures are listed in Table \ref{linepara}. The averaged C$^{18}$O (1-0) and $^{13}$CO (1-0) lines are optically thin for both $F_{M}$ and $F_{W}$. The excitation temperatures of C$^{18}$O (1-0) and $^{13}$CO (1-0) in $F_{M}$ are $\sim$15.0 K and 14.1 K, respectively. Being similar to the dust temperature, these temperatures suggest that C$^{18}$O and $^{13}$CO lines in $F_{M}$ are nearly thermally excited and that local thermodynamical equilibrium conditions hold. The excitation temperatures of the C$^{18}$O (1-0) and $^{13}$CO (1-0) line emissions in $F_{W}$, however, are $\sim$14.3 K and 14.5 K, respectively, which are lower than the dust temperature ($\sim$21 K), suggesting that C$^{18}$O (1-0) and $^{13}$CO (1-0) lines in $F_{W}$ are subthermally excited.

\section{Discussion}

\subsection{Massive filament formed due to Cloud-Cloud collision}

\begin{figure*}
\centering
\includegraphics[angle=90,scale=0.7]{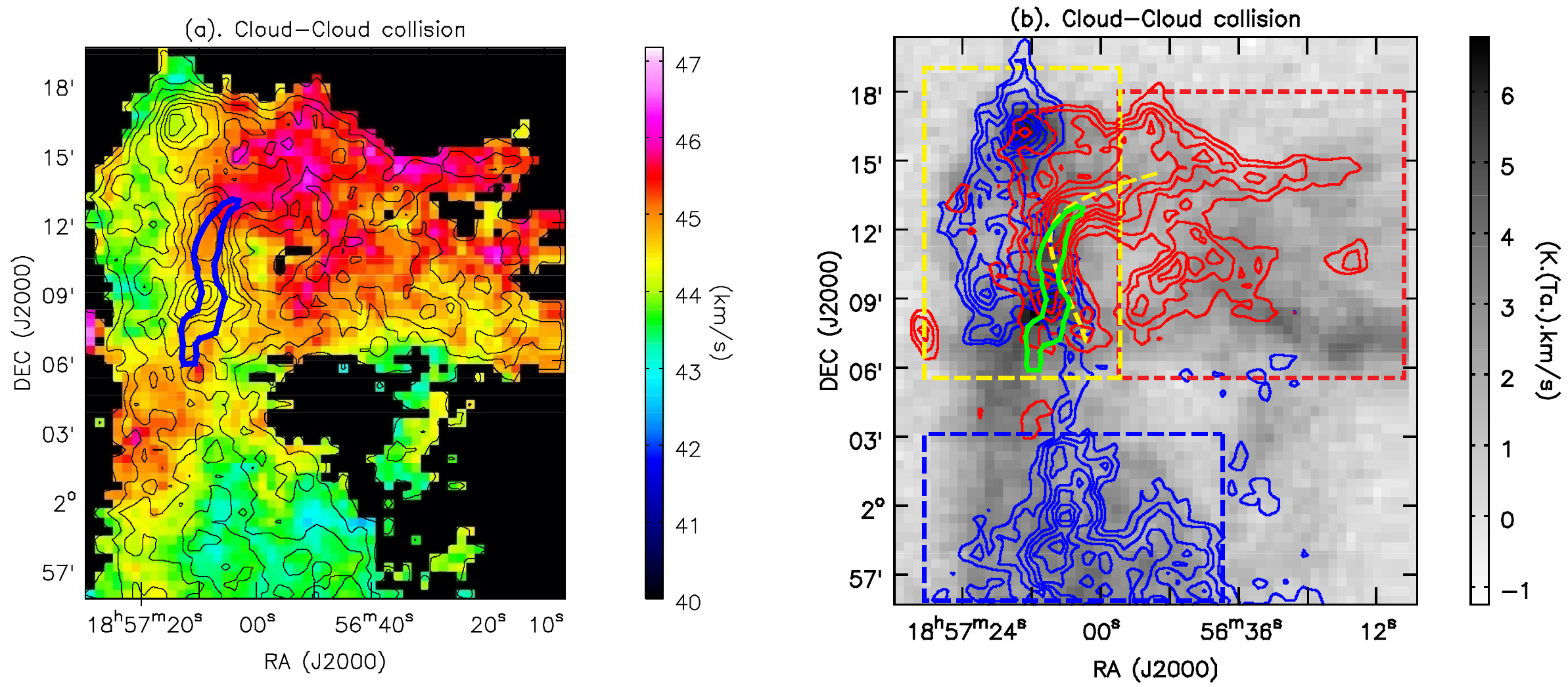}
\caption{(a). Moment 1 map of $^{13}$CO (1-0) emission. The black contours show integrated intensity in velocity channels 44 km~s$^{-1}$ and 45 km~s$^{-1}$. The contours are from 30\% to 90\% in steps of 10\% of the peak value (6.8 K~km~s$^{-1}$). The blue contour marks the main filament position. (b) The red (G035.39-west) and blue (G035.39-main) contours represent the cloud gas with redshifted velocities (46 - 47 km~s$^{-1}$) and blueshifted velocities (41 - 43 km~s$^{-1}$). The contour levels are from 30\% to 90\% in steps of 10\% of the peak values. The peak values for red and blue contours are 6.8 K~km~s$^{-1}$ and 4.0 K~km~s$^{-1}$, respectively. The green contour marks the main filament position. The red and blue dashed boxes mark the regions where redshifted emission or blueshifted emission dominates, respectively. The yellow dashed box marks the cloud-cloud collision area. The yellow dashed curve roughly shows the shell-like structure in the redshifted gas emission. \label{CCC} }
\end{figure*}

A large-scale, smooth velocity gradient of 0.4-0.8 km~s$^{-1}$~pc$^{-1}$ in the northern part of the main filament (\textit{$F_{M}$}) has been revealed in the $^{13}$CO, C$^{18}$O and N$_2$H$^+$ line emissions \citep{jim14,henshaw14}. Recently, \cite{soko17} mapped the whole \textit{$F_{M}$} in NH$_{3}$ lines and also found a smooth velocity gradient of $\sim$0.2 km~s$^{-1}$~pc$^{-1}$ across the whole filament as shown in panel (b) of Figure \ref{main}. Several scenarios have been proposed to explain these gradients, including cloud rotation, gas accretion along the filaments, global gravitational collapse, and unresolved sub-filament structures \citep{jim14}. A very promising scenario that could explain the presence of the smooth velocity gradient would involve the initial formation of sub-structures inside the turbulent molecular cloud, which interact with each other and may subsequently converge into each other as the cloud undergoes global gravitational collapse \citep{jim14}. From high-sensitivity and high-spectral-resolution molecular line (N$_{2}$H$^{+}$ and C$^{18}$O) observations, \cite{henshaw13} identify several velocity-coherent filaments inside \textit{$F_{M}$} and argue that \textit{$F_{M}$} formed via the collision of two relatively quiescent filaments moving at a relative velocity of $\sim$5 km~s$^{-1}$.

Based on our large-scale $^{13}$CO (1-0) map, we argue here that the velocity gradients and the collision of filaments inside \textit{$F_{M}$} are more likely caused by a large-scale ($\sim$10 pc) cloud-cloud collision. \textit{$F_{M}$} itself is also formed due to the large-scale cloud-cloud collision.

From the channel maps of $^{13}$CO (1-0) line emission (see the APPENDIX), we identify two velocity-coherent clouds (G035.39-main and G035.39-west) whose spatial distributions are distinctly different. Panel (a) of Figure \ref{CCC} presents the Moment 1 map of $^{13}$CO (1-0) line emission toward the G035.39 clouds. Blueshifted emission (G035.39-main) and redshifted emission (G035.39-west) are clearly separated in the northern part of the map, indicating two clouds in collision. The southern part of G035.39-main is not affected by cloud-cloud collision and thus shows no clear velocity gradient. \cite{hern15} argues that the large-scale velocity gradient in G035.39 is caused by cloud rotation. By inspecting the channel maps, however, the blueshifted emission gas and redshifted emission gas more likely belong to two different clouds. In addition, multiple velocity components in \textit{$F_{M}$} have been seen in high spectral resolution observations \citep{henshaw13}, which cannot be well explained by the cloud rotation scenario.

The integrated intensity maps of $^{13}$CO (1-0) for G035.39-west (46-47 km~s$^{-1}$) and G035.39-main (41-43 km~s$^{-1}$) are shown in the panel (b) of Figure \ref{CCC}. In the colliding area, the cloud gas with redshifted velocities is well separated spatially from the cloud gas with blueshifted velocities. \textit{$F_{M}$} is located in the interface layer, where the internal turbulence and the momentum exchange between the two colliding clouds may mix the gas distribution in both space and velocity and enhance the density therein. G035.39-west is curved as depicted by the yellow dashed line, suggesting that it is greatly compressed as it collides with G035.39-main. The widespread SiO emission in the northern part of \textit{$F_{M}$} discovered by \cite{jim10} may be the signs of (large-scale) shocks from the resulting compression. The emission peak of G035.39-west is in the interface, suggesting that majority of gas of the G035.39-west may have merged with the gas of G035.39-main. G035.39-west seems to have swung during collision forming a long tail in the west (as seen in the red dashed box). In addition, the northern (``N") part of \textit{$F_{M}$} appears to be more affected by the collision than the middle and southern parts (``M" and ``S") because the emission peak of G035.39-west is located close to the north end of \textit{$F_{M}$}. Moreover, the northern part of \textit{$F_{M}$} is found to have more redshifted velocities than the southern part, as seen from the Moment 1 map of $^{13}$CO (1-0) line emission. We suspect that the collision occurs from the north-west of \textit{$F_{M}$} and slows down as it propagates to the south, causing a velocity gradient along \textit{$F_{M}$}. Therefore, the previous findings of velocity gradients \citep{jim14,henshaw14,soko17} and multiple velocity components \citep{henshaw13} in \textit{$F_{M}$} can be explained by the mixed gas distribution from the two larger-scale colliding clouds.

The relative velocity between G035.39-west and G035.39-main is $\sim$5 km~s$^{-1}$, which is similar to that of the two colliding filaments suggested by \cite{henshaw13}. Considering projection effect, the collision velocity of the two clouds could be $\sim$10 km~s$^{-1}$ (for an inclination angle of 45$\arcdeg$), consistent with the collision speeds of GMCs in some simulations \citep{Inoue13,wu15,wu17}.  A cloud-cloud collision in G035.39 is also supported by [C{\sc ii}] observations \citep{bisbas18}.

\begin{figure*}
\centering
\includegraphics[angle=-90,scale=0.5]{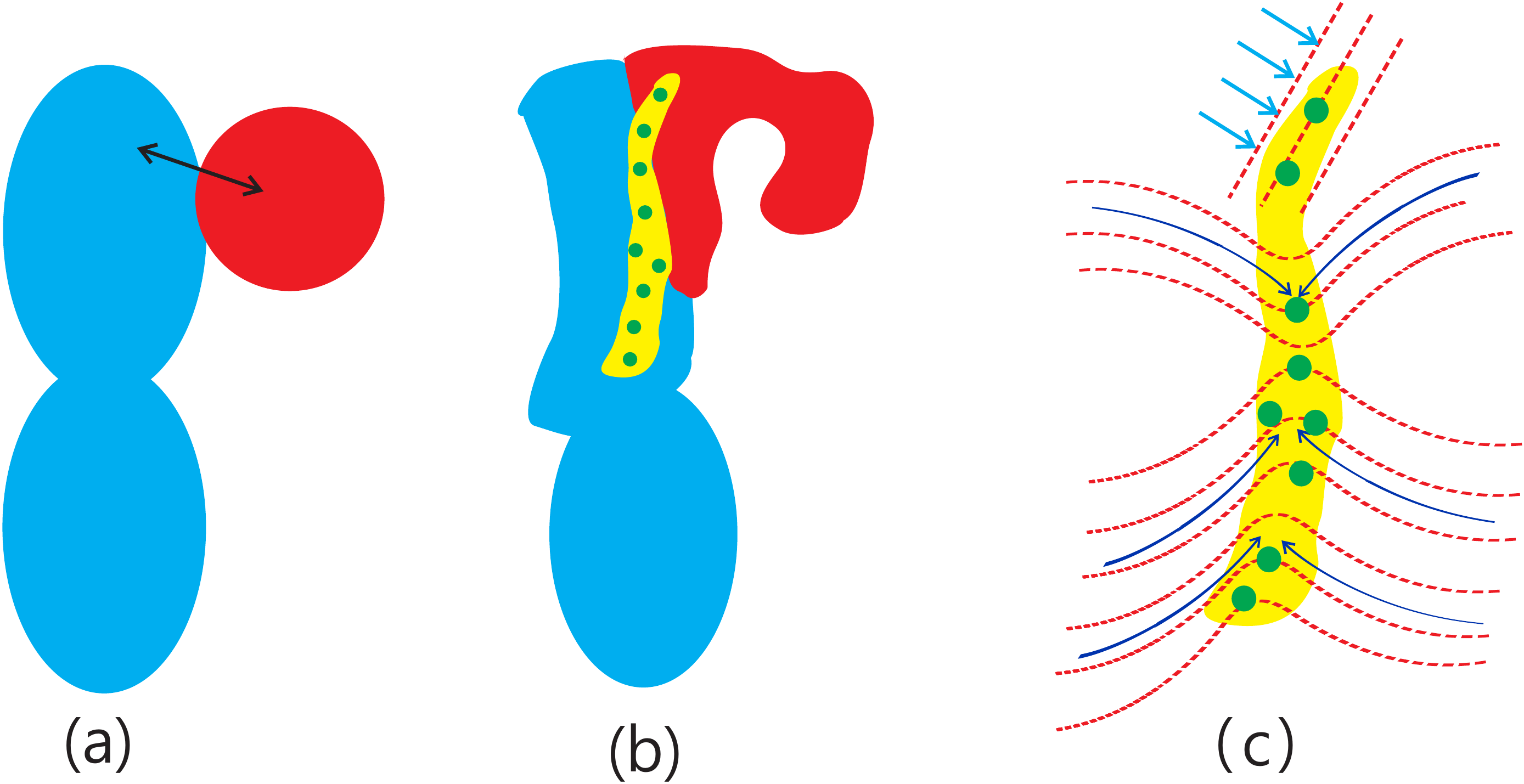}
\caption{Schematic illustration of the G035.39 clouds and magnetic fields: (a) Two clouds before collision. The blue one shows the original filamentary cloud, which fragments into two parts. The red one is colliding with the northern part of the filamentary cloud from the north-west direction. (b) Cloud-cloud collision enhances the density in the interface, where the massive filament \textit{$F_{M}$} is formed. \textit{$F_{M}$} is fragmented into dense clumps (green dots). The southern part of the filamentary cloud is not affected by the cloud-cloud collision and thus no dense structure is formed there. (c) Magnetic fields associated with \textit{$F_{M}$}. The red dashed lines show the magnetic field orientations. We note the magnetic field orientations offset from the filament are not well revealed by the present data due to the limited sensitivity to the lower density gas polarization signal. The blue arrows show the gas flow direction. \label{cart} }
\end{figure*}

A schematic illustration of the cloud-cloud collision is shown in panels (a) and (b) of Figure \ref{cart}. The smaller G035.39-west cloud collides with the northern part of G035.39-main in a north-west to south-east direction. The cloud-cloud collision enhances the density in the interface, where the massive filament \textit{$F_{M}$} has formed. The dynamical effect of the cloud-cloud collision may perturb \textit{$F_{M}$} and trigger its fragmentation. In contrast, the southern part (blue dashed box in panel (b) of Figure \ref{CCC}) of G035.39-main is not affected by the cloud-cloud collision and its density is not enhanced as seen from our $^{13}$CO map as well as infrared extinction maps \citep{kai13}. No dense clump (or new stars) has been formed there yet. Therefore, cloud-cloud collision can enhance density and shorten the local free-fall timescale for star formation.

\begin{figure}
\centering
\includegraphics[angle=0,scale=0.6]{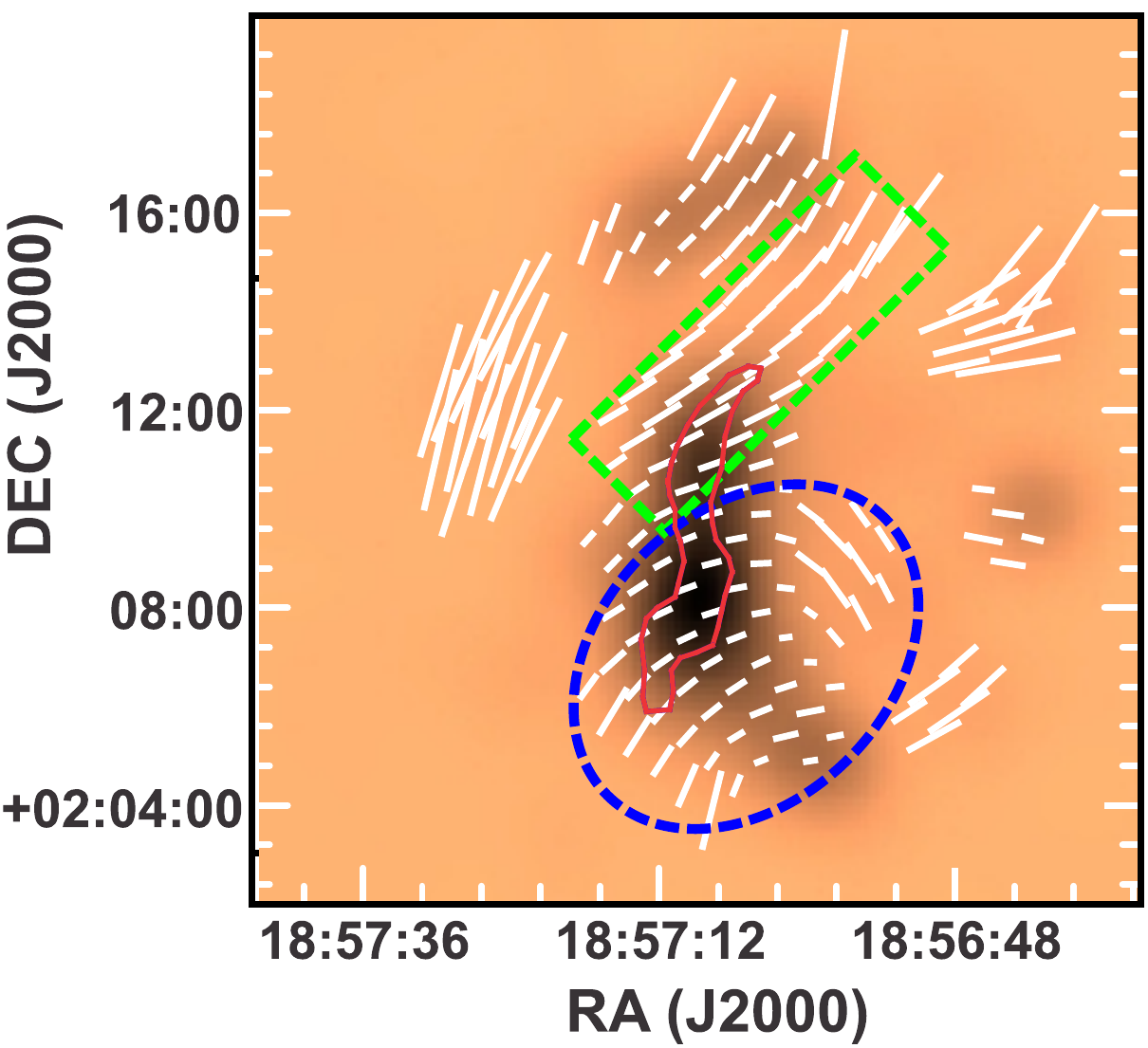}
\caption{Smoothed JCMT/POL-2 map of G035.39. The Q, U and I maps of G035.39 were smoothed with a beam size of 2$\arcmin$. The white orientations represent the smoothed magnetic field orientations. The red contour outline the massive filament \textit{$F_{M}$}. The green box and blue ellipse mark the regions showing different magnetic field geometry.  \label{smooth} }
\end{figure}

\subsection{The origin of magnetic field geometry surrounding the main filament}

As discussed in Section 3.1, the magnetic field orientations are roughly perpendicular to the major axis of \textit{$F_{M}$} along the central ridge and at the junctions with other filaments, while magnetic field orientations tend to be oblique in the lower density surroundings (see panel (d) in Figure \ref{main}). To explore the large-scale field geometry further, we convolve the Stokes Q, U, and I maps of POL-2 data with a 2$\arcmin$ beam and re-calculate the polarization angles from the smoothed maps. The smoothed magnetic field orientations are overlaid onto the similarly Stokes I image in Figure \ref{smooth}. The smoothed magnetic field orientations may indicate that the magnetic field is pinched around the middle and southern parts (marked by the blue dashed ellipse in Figure \ref{smooth}) of \textit{$F_{M}$}. The pinched magnetic field can be associated with the accretion flow around and along the filament in a globally collapsing cloud \citep[Li et al., 2018, in preparation; ][]{klass17,Gomez18}. Therefore, the magnetic fields in these regions seem to have been dragged by the collapsing gas flow which forms those dense structures \citep[Li et al., 2018, in preparation; ][]{lips15,klass17,Gomez18}.

As shown in Figure 1 in \cite{Gomez18}, the magnetic field lines can be dragged by the accretion flow. In low-density regions away from a filament, the gas flow direction is perpendicular to the filament and the dragged magnetic field must be mainly perpendicular to the filament as seen in the surroundings of other filamentary clouds \citep{chap11,cox16,sant16}. In the low-density regions surrounding the filament spine, the magnetic field is affected both by accretion flows onto and along the filament, and thus the magnetic field lines must also develop a component parallel to the filament and appear oblique. Accretion flows along the filament can compress the magnetic field at the filament spine and increases the perpendicular component as seen in simulations. This picture, illustrated in panel (c) of Figure \ref{cart}, can well explain the magnetic field geometry in the middle part of \textit{$F_{M}$}, where the projected magnetic field lines are mainly perpendicular to the filament along its densest regions and are oblique in less dense regions (see panel (d) in Figure \ref{main}).

As shown in panel (b) of Figure \ref{main}, the magnetic field orientations in the northern end of \textit{$F_{M}$} are nearly parallel to the filament. This pattern cannot be explained by gas flows along the dense filament because such flows would not increase the parallel component of the magnetic field but would rather increase the perpendicular component, leading to an ``U"-shaped magnetic geometry \citep[Li et al., 2018, in preparation; ][]{Gomez18}. As marked by the green dashed box in Figure \ref{smooth}, the smoothed magnetic field orientations in a large area close to the northern end of the \textit{$F_{M}$} are well aligned along the north-west to south-east direction. Indeed, such a nearly parallel pattern may be due to compression from the east (see panel (c) in Figure \ref{cart}). Since the northern part of \textit{$F_{M}$} is more affected by the cloud-cloud collision (see Section 4.1), the northern end of \textit{$F_{M}$} may be elongated and compressed by the two colliding clouds. The magnetic field at this location is well aligned with the elongated filament (see Figure \ref{cart}c).

\subsubsection{Pinched magnetic field surrounding clump ``c8"}

\begin{figure*}[tbh!]
\gridline{ \fig{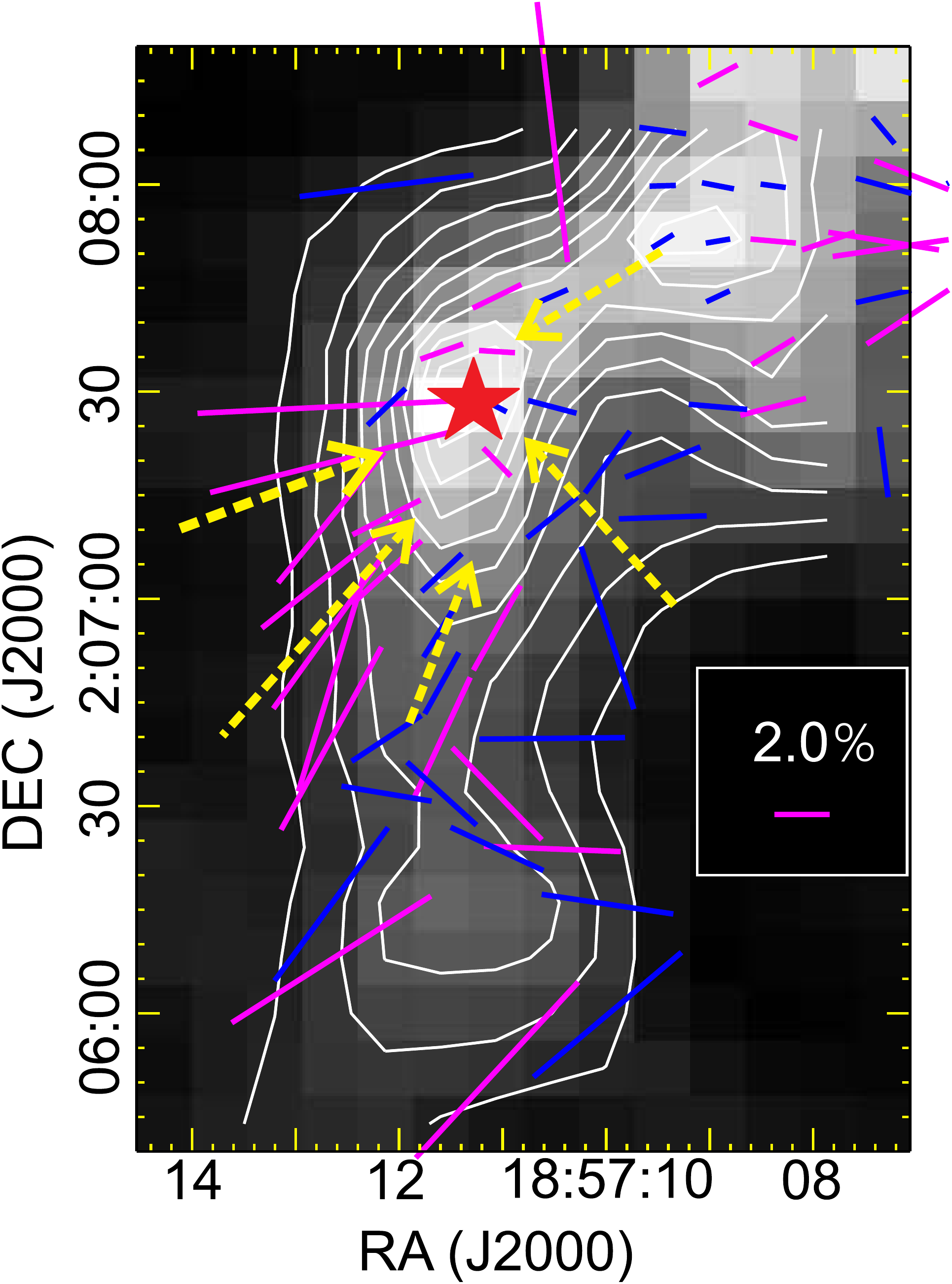}{0.29\textwidth}{(a). POL-2 map of clump ``c8"}
            \fig{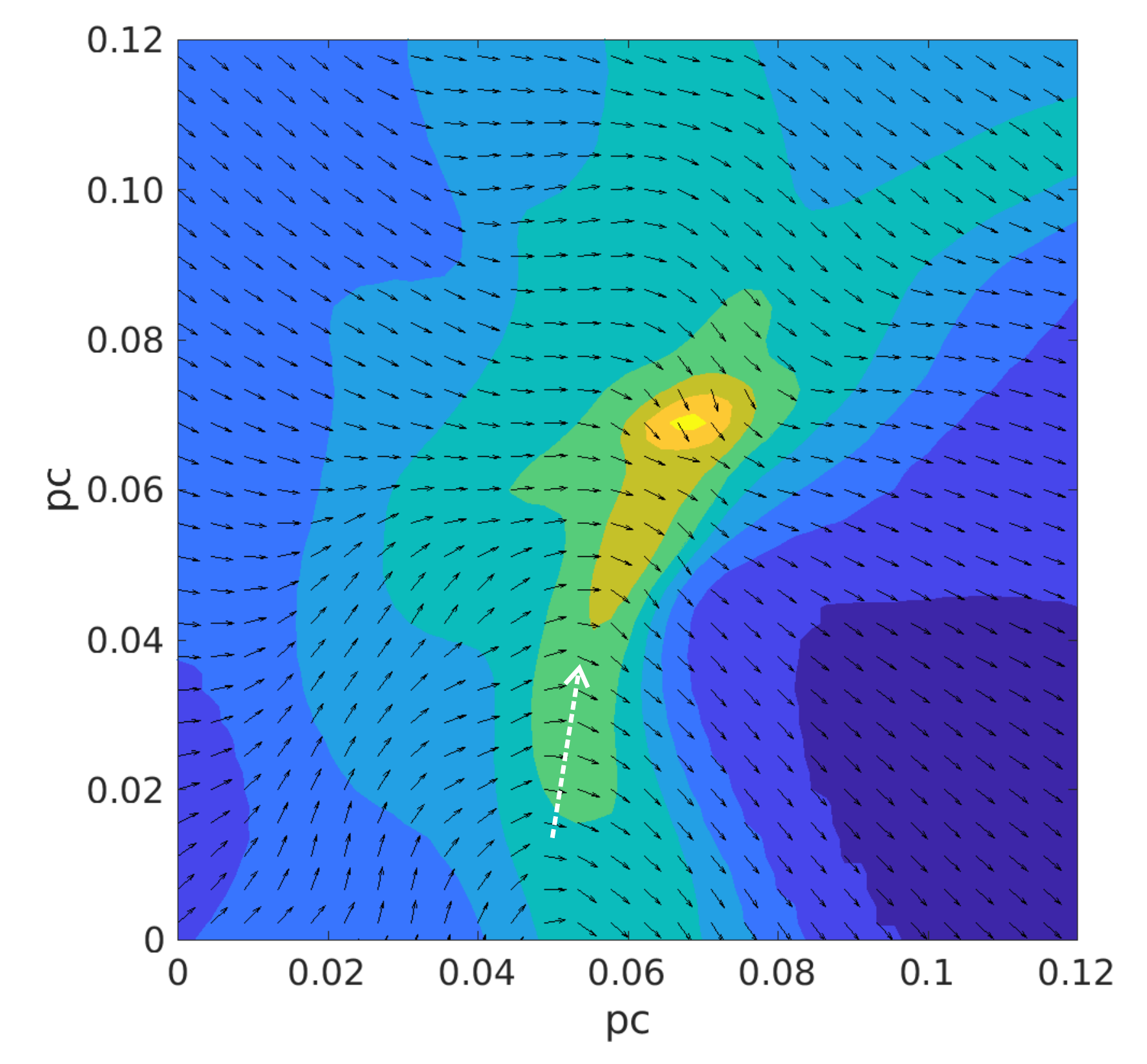}{0.38\textwidth}{(b). Magnetic field streamlines surrounding an accreting core in simulations}
            \fig{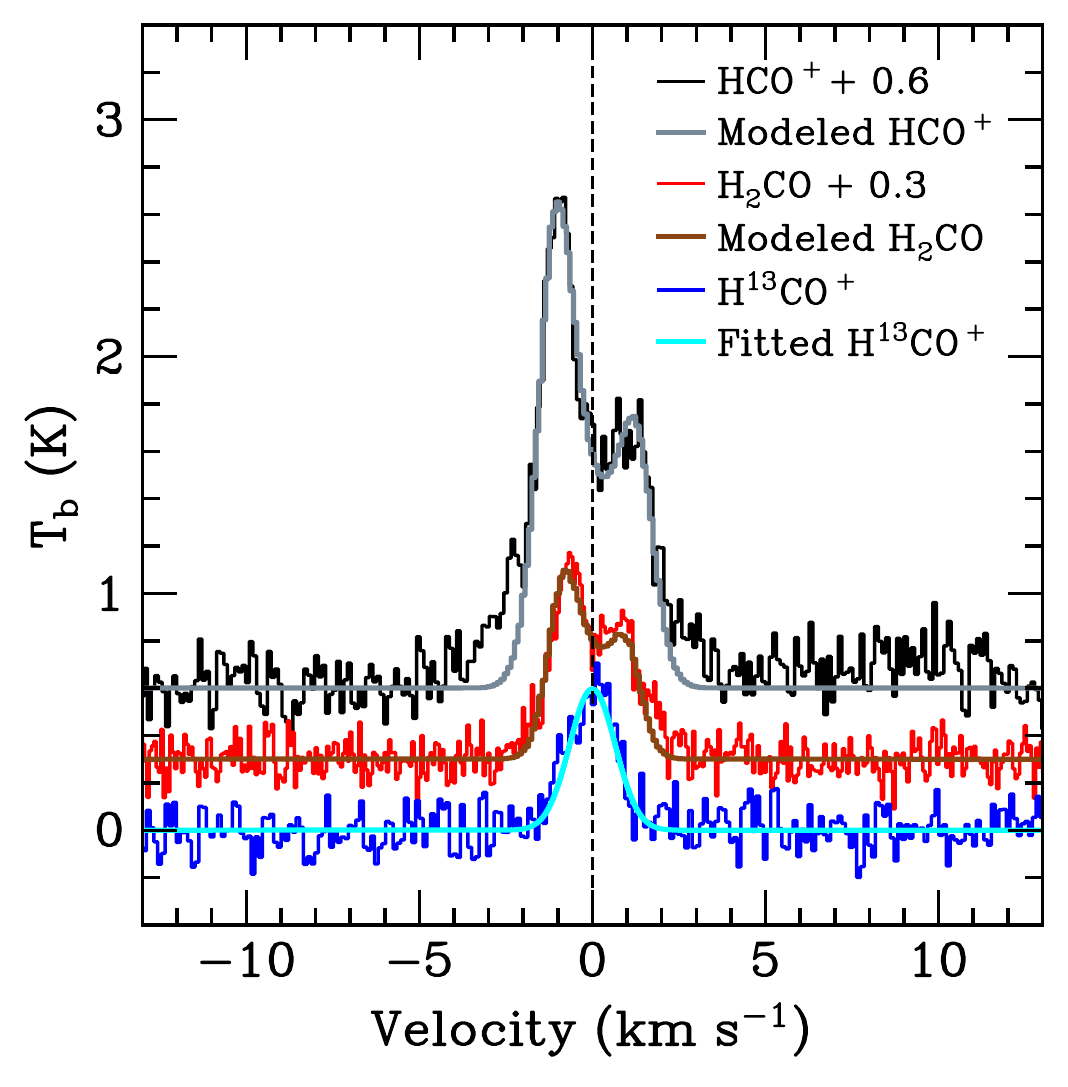}{0.33\textwidth}{(c). Spectra at the central position of ``c8"}
            }
\caption{\small (a) POL-2 map of the global collapsing clump ``c8". The background is Stokes I at 850 $\micron$. The pixel size is 8 arcsec. The magenta orientations are measurements with $SNR>$3 for P. The blue orientations are measurements with $2<SNR<3$ for P. The cutoff for Stokes I is $SNR>30$. The length of the orientations represent the corresponding polarization fraction. The contours are from 10\% to 90\% in steps of 10\% of the peak value of 375 mJy/beam. The emission peak of the massive starless clump candidate is marked by a star symbol. The yellow dashed lines show the suggested gas flow directions. (b) Magnetic field surrounding an accreting core projected on the column density map in simulations (Li et al., 2018, in preparation). The core is still accumulating gas along the filament. The white dashed arrow shows the accreting direction. (c) Spectra from line observations with KVN 21-m telescope. HCO$^{+}$ (1-0), H$_2$CO ($2_{1,2}-1_{1,1}$), and H$^{13}$CO$^{+}$ (1-0) are shown in black, red, and blue, respectively. The H$^{13}$CO$^{+}$ (1-0) line has been fitted with a Gaussian function (blue line). The dashed vertical line represents the systemic velocity of 44.9 km~s$^{-1}$. The best fits from RATRAN models toward HCO$^{+}$ (1-0) and H$_2$CO ($2_{1,2}-1_{1,1}$) are also shown overlaid on the observed spectra. \label{core}}
\end{figure*}

The magnetic field surrounding clump ``c8" is shown in panel (a) of Figure \ref{core}. The magnetic field orientations associated with ``c8" are likely pinched. The pinched magnetic field may hint at gas inflows toward the center of ``c8". The yellow dashed arrows in the panel indicate the possible gas flow directions. Interestingly, the magnetic field orientations are roughly parallel to the suggested gas flow directions, a behaviour seen also in simulations \citep[Li et al., 2018, in preparation; ][]{klass17,Gomez18}. In addition, the magnetic field orientations close to the center become more perpendicular to the major axis of the local filament, suggesting that magnetic fields therein have been compressed by accretion flows and become more perpendicular as a result \citep[Li et al., 2018, in preparation; ][]{klass17,Gomez18}. Indeed, the magnetic fields surrounding ``c8" hint at a ``U"-shaped geometry caused by accretion flows \citep[Li et al., 2018, in preparation; ][]{Gomez18}. The coarse resolution of our POL-2 observations, however, cannot resolve the field geometry of the clumps. In addition, there are only a handful of high signal-to-noise orientations, too few to well constrain the field geometry. Future higher sensitivity and resolution polarization observations are needed to investigate the field geometry in greater detail.

In panel (b) of Figure \ref{core}, we show the magnetic field surrounding an accretion core from a simulation of an IRDC \citep[Li et al., 2018, in preparation; ][]{lips15}. This image is taken from an ideal MHD turbulence simulation driven at thermal Mach number of 10 and Alfven Mach number of 1. The image is selected at half of the freefall time of the simulation. The core (bright yellow region) is accreting gas along the filament. The white dashed arrow shows the accretion direction. The magnetic field has been twisted along the accretion direction. Close to the densest part of the core, the magnetic field is compressed and is roughly perpendicular to the filament. The accretion significantly increases the perpendicular component of the magnetic field at the filament spine. Although ``c8" is much more massive and larger than the simulated core, the magnetic field surrounding ``c8" shows similar geometry to that associated with the simulated core, suggesting that magnetic field surrounding ``c8" has been similarly pinched by gas inflow along the filament.

Using ALMA, \cite{henshaw17} resolved clump ``c3" into a network of narrow ($\sim0.028\pm0.005$ pc in width) sub-filaments which contain 28 compact cores. Those cores may be still accreting gas along those sub-filaments. Therefore, we suspect that the magnetic fields surrounding those cores are also pinched due to the accretion, as seen in simulations (see panel (b) of Figure \ref{core}). Future polarization observations with ALMA will shed light on the magnetic field geometry surrounding these cores.

\subsection{Gravitational Stability of the main filament}

Most stability analysis of massive filaments suggest that the magnetic field is important, but a thorough analysis has been elusive given the difficulties in observing magnetic fields \citep[e.g.,][]{jack10,cont16,henshaw16,lu18}. In this section, we investigate the gravitational stability of the main filament $F_{M}$ by taking into account thermal pressure, turbulence, and magnetic fields.

The critical line mass for the global gravitational stability of an isothermal filament supported by thermal pressure and turbulence is \citep{ostri64,pattle17}:
\begin{equation}
(\frac{M}{L})_{crit}=\frac{2\sigma_{3D}^2}{G}
\end{equation}
where G is the gravitational constant. Assuming that the velocity dispersion is isotropic, the three-dimensional velocity dispersion is:
\begin{equation}
\sigma_{3D}=\sqrt{3(\sigma_{NT}^2+c_{s}^2)}
\end{equation}
We obtain a mean non-thermal velocity dispersion ($\sigma_{NT}$) of $\sim$0.4 km~s$^{-1}$ from the line widths of the NH$_{3}$ (1,1) line \citep{soko17}. The 1D thermal velocity dispersion (or sound speed) is:
\begin{equation}
c_{s}=\sqrt{\frac{kT_{k}}{\mu m_{H}}}
\end{equation}
$\mu=2.37$ is the mean molecular weight per ``free particle" (H$_2$ and
He, the number of metal particles is negligible). $c_{s}$ is $\sim$0.23 km~s$^{-1}$ for a mean kinetic temperature ($T_{\textrm{k}}$) of 15 K. Here we assume that $T_{\textrm{k}}$ equals $T_{\textbf{d}}$ under the local thermodynamic equilibrium (LTE) conditions. Therefore, the $\sigma_{3D}$ and $(\frac{M}{L})_{crit}$ are 0.80 km~s$^{-1}$ and $\sim$296 M$_{\sun}$~pc$^{-1}$, respectively. The $(\frac{M}{L})_{crit}$ is smaller than the measured line mass ($\sim$410 M$_{\sun}$~pc$^{-1}$), suggesting that the filament cannot be supported against collapse only by turbulent gas pressure in the absence of magnetic fields.

In contrast, the $F_{W}$ filament is as turbulent as the $F_{M}$ but is much less dense. The critical line mass for $F_{W}$ is comparable to that of $F_{M}$, while its line mass is only $\sim$100 M$_{\sun}$~pc$^{-1}$. Therefore, turbulent motions in $F_{W}$ can dominate over gravity in $F_{M}$ and may further stretch $F_{W}$.

The criterion for filament stability with support from magnetic fields can be estimated as \citep{ostri64,Fiege00,pattle17}:
\begin{equation}
(\frac{M}{L})_{crit,mag}=(\frac{M}{L})_{crit}(1-\frac{\mathcal{M_B}}{|\mathcal{W}|})^{-1},
\end{equation}
where $\mathcal{M_B}$ is the magnetic energy per unit length, and $|\mathcal{W}|=(\frac{M}{L})G\approx4.6\times10^{26}$ erg~cm$^{-1}$ is the gravitational energy per unit length. $\mathcal{M_B}$ may be either positive or negative, depending on whether a poloidal or the toroidal field, respectively, dominates the overall magnetic energy \citep{Fiege00}. If a poloidal field dominates, the factor $(1-\frac{\mathcal{M_B}}{|\mathcal{W}|})^{-1}$ is larger than 1 \citep{Fiege00}. Therefore, a poloidal field helps to support the cloud radially against self-gravity by increasing the critical mass per unit length \citep{Fiege00}. In contrast, the factor $(1-\frac{\mathcal{M_B}}{|\mathcal{W}|})^{-1}$ is smaller than 1 \citep{Fiege00} for a toroidal field. A toroidal field works with gravity in squeezing the cloud by reducing the critical mass per unit length \citep{Fiege00}.

The total magnetic field strength ($B_{tot}$) can be estimated using the Davis-Chandrasekhar-Fermi (DCF) method \citep{davis51,chandr53}: \begin{equation}
B_{tot}=1.3B_{pos}=1.3Q'\sqrt{4\pi\rho}\frac{\sigma_{NT}}{\sigma_{\theta}},
\end{equation}
where $B_{pos}$ is the POS magnetic field strength, 1.3 is a factor considering projection effects, $Q'$ is a factor of order unity accounting for variations in field strength on scales smaller than the beam \citep{crut04}, $\rho=\mu_g m_{H}n_{H_2}$ is the gas density, and $\sigma_{\theta}$ is the dispersion in polarization position angles. Here $Q'$ is taken as 0.5 \citep{ostr01}.

As seen in Figure \ref{main}, the magnetic field orientations in the middle part of $F_{M}$ are well ordered and uniform with their orientations roughly perpendicular to the major axis of the filament. In addition, the magnetic field orientations tend to be more perpendicular toward the denser regions (see panel d of Figure \ref{main}). In contrast, the magnetic field orientations in the northern and southern ends are more widely dispersed in direction. We only estimated $\sigma_{\theta}$ in the middle part of the main filament where the Stokes I intensity at 850 $\micron$ is above 100 mJy~beam$^{-1}$. As shown in panel (a) of Figure \ref{stats}, from a Gaussian fit to the orientation angles, the measured $\sigma_{\theta}$ is $\sim$17$\arcdeg$. We correct the angular dispersion $\sigma_{\theta}$ by subtracting the measurement uncertainty ($\delta_{\theta}\sim9\arcdeg$) with $\sqrt{\sigma_{\theta}^2-\delta_{\theta}^2}$. The corrected $\sigma_{\theta}$ is $\sim$15$\arcdeg$, which is smaller than the maximum value at which the standard DCF method can be safely applied \citep[$\leq25\arcdeg$; ][]{heit01}. Taking $\sigma_{\theta}$ as $\sim$15$\arcdeg$, we obtain a POS magnetic field strength ($B_{pos}$) of 50 $\mu G$ and hence a total magnetic field strength ($B_{tot}$) of 65 $\mu G$.

The estimated magnetic field strength should be treated as an upper limit because: (1) $\sigma_{\theta}$ is estimated from the densest region of $F_{M}$ with uniform magnetic field orientations. The magnetic field orientations in other parts of $F_{M}$ are more widely dispersed and should have larger $\sigma_{\theta}$. Therefore, $\sigma_{\theta}$ used in the above calculations with the DCF method is a lower-limit. (2) $\sigma_{NT}$ is estimated from the mean line width of NH$_{3}$ (1,1) \citep{soko17}. The GBT beam at NH$_{3}$ (1,1) line frequency is 32$\arcsec$, larger than the 14.1$\arcsec$ beam of SCUBA-2/POL-2 at 850 $\micron$. Therefore, some uncertainties remain in $\sigma_{NT}$ for the estimation of magnetic field strength using the DCF method at this scale.

\cite{crut10} obtained an empirical relation for a maximum field strength ($B_{max}$) versus density from Zeeman observations:
\begin{equation}
B_{max}\simeq0.22(\frac{n_{H}}{10^5~cm^{-3}})^{0.65}~mG~~(n_{H}>300~cm^{-3})
\end{equation}
For a mean $n_{H}$ of $\sim1.5\times10^4$ cm$^{-3}$ in $F_{M}$, the maximum field strength $B_{max}$ estimated from the empirical relation is 64 $\mu G$, i.e., similar to the value (65 $\mu G$) derived from the DCF method. Therefore, 65 $\mu G$ may represent an upper limit for the mean magnetic field strength of the main filament $F_{M}$.

Although lacking of accurate determination of magnetic field strength, we can use the upper limit of 65 $\mu G$ to evaluate the importance of magnetic field in the gravitational stability of the main filament.

The corresponding Alf\'{v}enic velocity for a magnetic field strength of 65 $\mu G$ is:
\begin{equation}
\sigma_{A}=\frac{B_{tot}}{\sqrt{4\pi\rho}},
\end{equation}
where $\sigma_{A}\approx1.0$ km~s$^{-1}$ for $B_{tot}$=65 $\mu G$ and a mean volume density of $7.3\times10^3$ cm$^{-3}$. The Alf\'{v}en Mach number is \begin{equation}
\mathcal{M_A}=\sqrt{3}\sigma_{NT}/\sigma_{A}.
\end{equation}
We derived $\mathcal{M_A}\approx0.7$, suggesting that the turbulent motions may be sub-Alfv\'{e}nic in the main filament. The total magnetic energy ($E_{B}$) is \citep{pattle17}:
\begin{equation}
E_{B}=\frac{B_{tot}^2V}{2\mu_0}
\end{equation}
where $\mu_{0}$ is the permeability of free space. Therefore, the magnetic energy per unit length is
\begin{equation}
|\mathcal{M_B}|=\frac{E_{B}}{L}
\end{equation}
Considering the volume ($V$=5.4~pc$^3$) and length ($L$=6.8~pc) of $F_{M}$, $E_{B}$ and $|\mathcal{M_B}|$ are $\sim2.7\times10^{46}$ erg and $\sim1.3\times10^{26}$ erg~cm$^{-1}$, respectively. Therefore, if a poloidal field component dominates, it will increase the critical line mass by a factor of $(1-\frac{1.3\times10^{26}}{4.6\times10^{26}})^{-1}\sim1.39$. In this case, the critical line mass, taking into account the additional support from magnetic fields, will become $\sim$411 M$_{\sun}$~pc$^{-1}$, which is very similar to the measured value ($\sim$410 M$_{\sun}$~pc$^{-1}$). If, however, a toroidal field component dominates, it will decrease the critical line mass by a factor of $(1-\frac{-1.3\times10^{26}}{4.6\times10^{26}})^{-1}\sim0.78$. If so, the critical line mass will become $\sim$231 M$_{\sun}$~pc$^{-1}$, much smaller than the measured value ($\sim$410 M$_{\sun}$~pc$^{-1}$).

Judging from panels (a) and (b) in Figure \ref{main}, the northern part (``N") of $F_{M}$ has magnetic field orientations parallel to the major axis, indicating that the magnetic field therein is likely poloidal. Therefore, the northern part may be stable with additional support from magnetic fields.

In contrast, the middle part of $F_{M}$ is dominated by a magnetic field whose orientation is perpendicular to the major axis, resembling the projection of a toroidal magnetic field wrapping around the filament. If so, the middle part may become more unstable. Alternatively, the field can also just simply go straight through the filament, providing no support against gravitational collapse. Therefore, the middle part is likely unstable and may further fragment or collapse. Indeed, clump ``c3" in the middle part already contains plenty of substructures (sub-filaments and cores) detected in ALMA observations \citep{henshaw17}.

The southern part of $F_{M}$ is more complicated. In contrast to the middle part, the magnetic field orientations in the southern part are more parallel to the major axis. We note, however, that the magnetic field orientations tend to be more perpendicular to the major axis in the densest region of the southern part. Therefore, the magnetic field in the southern part may contain comparable toroidal and poloidal components. The critical line mass considering magnetic field support will not deviate too much from that without magnetic field support. Hence, the southern part may be unstable and may fragment or collapse.

\subsection{Gravitational Stability of dense clumps}

In this section, we investigate the gravitational stability of dense clumps from virial analysis by taking into account the support from thermal pressure, turbulence and magnetic fields.

If we only consider support from thermal pressure and turbulence, the virial masses ($M_{vir}$) of the clumps, assuming a uniform density profile, are \citep{bert92,pillai11,sanh17}:
\begin{equation}
M_{vir}=\frac{5R_{eff}}{G}(\sigma_{NT}^2+c_{s}^2)
\end{equation}
The virial masses calculated are presented in Table \ref{clumppara}. Three clumps (``c1", ``c2" and ``c5") have virial masses that are two to three times larger than their clump masses and hence may be gravitationally unbound, suggesting that ``turbulent" gas motions in the clumps provide enough support against self-gravity. The other clumps have virial masses smaller than their clump masses, suggesting that they are bound and unstable without additional support from magnetic fields.

\cite{henshaw16} also suggested that the dense cores revealed in high-resolution interferometric observations are susceptible to gravitational collapse without additional support from magnetic fields. Therefore, it is important to evaluate the importance of magnetic fields in the gravitational stability of the dense clumps. Our SCUBA-2/POL-2 observations, however, do not resolve the magnetic fields surrounding those clumps, and thus we do not have estimation of their magnetic field strengths from observations. Instead, we estimate the magnetic field strengths with the empirical relation from \cite{crut10} and \cite{lips15}.

Based on their MHD simulation results, \cite{lips15} suggested that the average field strength ($B_{clump}$) in molecular clumps in the interstellar medium is:
\begin{equation}
B_{clump}\simeq42(\frac{n_{H}}{10^4~cm^{-3}})^{0.65}~\mu G
\end{equation}

Using equation (18), we estimated the total magnetic field strength $B_{clump}$ for clumps. The $B_{clump}$ and Alfv\'{e}nic speed $\sigma_{A}$ of clumps are listed in Table \ref{clumppara}. The $B_{clump}$ values range from $\sim$56 $\mu$G to 219 $\mu$G. The mean Alf\'{v}en Mach number of clumps is $\sim$0.75, suggesting that the magnetic field may play a role as important as turbulence in supporting clumps against gravity. To investigate the gravitational stability of those dense clumps, we estimated the virial masses ($M_{vir}^{B}$) of the clumps considering thermal, turbulent, and magnetic pressures and assuming a uniform density profile \citep{bert92,pillai11,sanh17}:
\begin{equation}
M_{vir}^{B}=\frac{5R_{eff}}{G}(\sigma_{NT}^2+C_{s}^2+\frac{\sigma_{A}^2}{6})
\end{equation}
$M_{vir}^{B}$ are also presented in Table \ref{clumppara}. Three clumps (``c1", ``c2" and ``c5") have virial masses two to three times larger than their clump masses and hence may be gravitationally unbound, suggesting that ``turbulent" gas motions and magnetic fields in them provide significant support against self-gravity. The most two massive clumps (``c3" and ``c8"), however, have clump masses larger than its virial masses, suggesting that they will collapse and fragment. The other clumps have virial masses comparable to their clump masses, suggesting that they are close to virial equilibrium with additional support from magnetic fields.

Clump ``c8" is particularly interesting because it is not visible at Herschel/PACS 70 $\micron$ and 160 $\micron$ bands as well as Spitzer/MIPS 24 $\micron$ band, indicating that it is very cold and maybe starless. The physical parameters (e.g., mass, density, size) of ``c8" are similar to other Galactic massive starless clumps discovered in large surveys \citep[e.g.,][]{Guzm15,traf15,cont17,yuan17}. As noted earlier (Section 4.2.1), the magnetic field surrounding ``c8" is pinched, hinting at gas inflow along the filament. The virial parameter ($\alpha_{vir}$) of ``c8" is $\alpha_{vir}=M_{vir}/M_{clump}\leq0.6$ even if we consider additional support from magnetic fields, suggesting that ``c8" is undergoing gravitational collapse.

Figure \ref{core} shows an evidence of the collapse of ``c8" from the asymmetric ``blue-skewed profiles" of optically thick lines (HCO$^{+}$ (1-0) and H$_2$CO ($2_{1,2}-1_{1,1}$) from KVN observations. The systemic velocity of ``c8" is 44.9 km~s$^{-1}$, which is determined from Gaussian fitting to the single-peaked H$^{13}$CO$^{+}$ (1-0) line. In contrast, HCO$^{+}$ (1-0) and H$_2$CO ($2_{1,2}-1_{1,1}$) show double-peaked emission with the blueshifted peak stronger than the redshifted one, a typical ``blue-skewed profiles" for infall signature \citep{zhou93}. Such ``blue-skewed profile" of optically thick lines is commonly seen in surveys toward massive clumps \citep{wu03,wu07,full05,liu16b,jin16}, which can be interpreted as an evidence for the global collapse of massive clumps \citep{pere13,liu13b}. We highlight that, to our knowledge, ``c8" could be the first discovered massive starless clump candidate exhibiting this characteristic infall profile.

We model the HCO$^{+}$ (1-0) and H$_2$CO ($2_{1,2}-1_{1,1}$) lines using RATRAN following \cite{pere13} and \cite{yuan18}. For the modelling, a power-law density profile ($\rho\propto r^{-1.5}$) is assumed and the kinetic temperature is set to be 13 K. We have tried a grid of models by varying molecular abundances, infall velocities, and velocity dispersions. The resulting infall velocity inferred from the best models for HCO$^{+}$ (1-0) is 0.32$\pm$0.04 km~s$^{-1}$ while the resulting infall velocity derived from the best models for H$_2$CO ($2_{1,2}-1_{1,1}$) is 0.20$\pm$0.10 km~s$^{-1}$. Though the values are arguably the same within the uncertainties, the infall velocity traced by H$_2$CO ($2_{1,2}-1_{1,1}$) is smaller than that traced by HCO$^{+}$ (1-0). Since the effective excitation density (1.5$\times10^5$ cm$^{-1}$) of H$_2$CO ($2_{1,2}-1_{1,1}$) at 10 K is much larger than that (9.5$\times10^2$ cm$^{-1}$) of HCO$^{+}$ (1-0) \citep{shir15}, H$_2$CO ($2_{1,2}-1_{1,1}$) should trace denser, inner regions of the clump than HCO$^{+}$ (1-0). Therefore, the gas inflow indicated by H$_2$CO ($2_{1,2}-1_{1,1}$) and HCO$^{+}$ (1-0) seems to be decelerated from the outer part to inner part. The decelerated inflow may be caused by the enhanced magnetic field strength near the clump center, which will help resist gravity. Considering the uncertainties in the infall velocities, future work is needed.

Assuming a power-law density profile ($\rho\propto r^{-1.5}$), the mass enclosed in $r_{o}$ is
\begin{equation}
M=\int_0^{r_o}4\pi r^2\rho_{o}(\frac{r}{r_o})^{-1.5}dr=\frac{4\pi}{1.5}r_{o}^3\rho_{o}
\end{equation}
where $r_{o}$ is the outer radius and $\rho_{o}$ is the density at $r_{o}$. Therefore, the mass inflow rate at $r_{o}$ can be estimated using:
\begin{equation}
\dot{M}_{in}=4\pi r_{o}^2\rho_{o} v_{in}=1.5Mv_{in}/r_{o}
\end{equation}
We take the total clump mass ($\sim$200 M$_{\sun}$) for M and clump radius (0.28 pc) for r$_{o}$. We assume an infall velocity $v_{in}$ of 0.32 km~s$^{-1}$ obtained from the HCO$^{+}$ (1-0) measurement because the mean volume density of ``c8" is closer to the critical density of HCO$^{+}$ (1-0). The inferred mass inflow rate ($\dot{M}_{in}$) is thus $\sim4\times10^{-4}$ M$_{\sun}$yr$^{-1}$. This mass inflow rate is consistent with those measured in other high-mass star forming clumps \citep{wu09,wu14,sanh10,liu11a,liu11b,liu13a,liu13b,liu16a,ren12,pere13,qin16,yuan18}. The clump mass of ``c8" also exceeds the empirical threshold ($M>870M_{\sun}(r/pc)^{1.33}$) for high-mass star forming clumps discovered by \cite{kauf10}. In addition, the clump mass of ``c8" is also comparable to the masses of high-mass starless clumps with similar radii catalogued by \cite{yuan17}. All those indicate that ``c8" has the potential to form high-mass stars.

The collapsing massive starless clump candidate ``c8" may represent the very initial conditions for high-mass star formation and deserves more detailed studies at higher angular resolution. Indeed, searching for the existence or absence of high-mass prestellar cores, as it has been done in other massive starless clump candidates \citep{Tan13,beut13,sanh13,sanh17,liu17,cont18}, is of great importance given that in ``c8" the magnetic field and infall speed at large scales are now both known, unlike for the other studies.

\section{Summary}

We have studied the magnetic fields projected on the plane of the sky in the massive IRDC G035.39-00.33 from the JCMT/POL-2 polarization observations at 850 $\micron$ and the large scale kinematics from  various molecular line observations. Our main findings are summarized below.

(1) From the deep JCMT/POL-2 observations, we identified a network of elongated structures covering a broad range of densities. The most massive filament (F$_{M}$) has a length of $\sim$6.8 pc, a mass of $\sim$2800 M$_{\sun}$, and a line mass of $\sim$410 M$_{\sun}$~pc$^{-1}$. The other fainter elongated structures have comparable lengths but are much less dense. A long elongated structure (F$_{W}$) having a length similar to F$_{M}$ is connected to the northern end of F$_{M}$. F$_{W}$ is about four times less massive and less dense than F$_{M}$.

(2) The orientations of the magnetic fields in the two less dense tails of F$_{M}$ and some other less dense elongated structures (e.g., F$_{W}$) tend to be parallel to the major axes of their respective skeletons. In contrast, magnetic fields in the densest regions of the middle part of F$_{M}$ and some nodes at its junctions with other elongated structures (\textit{$F_{SW}$}, \textit{$F_{E}$}, and \textit{$F_{NE}$}) are more perpendicular to the major axis.

(3) We claim that the massive filament F$_{M}$ forms at the interface of two colliding clouds. The large-scale velocity gradient and multiple velocity components in F$_{M}$ discovered in previous works can be now explained by the mixed gas distribution from these two colliding clouds. The northern end of F$_{M}$ is more compressed by the cloud-cloud collision and the magnetic fields therein are also compressed and aligned along the filament.

(4) F$_{M}$ is unstable against gravity if we only consider internal support from thermal pressure and turbulence. The magnetic field orientations suggest that the northern part of F$_{M}$ may be dominated by a poloidal magnetic field component, which may provide additional support against gravity by increasing the effective critical mass per unit length. In contrast, the middle part of F$_{M}$ may be dominated by a toroidal field component, which reduces the effective critical mass per unit length and makes the filament more unstable. The southern part of F$_{M}$ is also unstable, even considering support from the magnetic field.

(5) Nine clumps with masses ranging from 16 M$_{\sun}$ to 219 M$_{\sun}$ are identified along the main filament F$_{M}$. The gravitational stability of the clumps is evaluated from a virial analysis considering internal support from thermal pressure, turbulence and magnetic fields. Three clumps (``c1", ``c2", and ``c5") have virial masses much larger than their clump masses and hence are gravitationally unbound. The two most massive clumps (``c3" and ``c8"), however, have clump masses larger than their virial masses even if the magnetic field support is considered, suggesting that they will collapse and fragment. The other clumps have virial masses comparable to their clump masses, suggesting that they are close to virial equilibrium, with additional support from magnetic fields.

(6) We discovered a massive ($\sim$200 M$_{\sun}$), collapsing starless clump candidate, ``c8". This clump has a clump mass about two times larger than its virial mass, suggesting it will collapse and fragment. The magnetic field surrounding ``c8" is pinched, likely due to the accretion flow along its host filament. HCO$^{+}$ (1-0) and H$_2$CO ($2_{1,2}-1_{1,1}$) spectra toward ``c8" show a clear infall signature, i.e., the ``blue-skewed profile". The infall velocities inferred from HCO$^{+}$ (1-0) and H$_2$CO ($2_{1,2}-1_{1,1}$) are 0.32$\pm$0.04 km~s$^{-1}$ and 0.20$\pm$0.10 km~s$^{-1}$, respectively. The mass inflow rate is $\sim4\times10^{-4}$ M$_{\sun}$~yr$^{-1}$. As this rate is consistent with those measured in other high-mass star forming clumps, ``c8" likely has potential ability to form high-mass stars. Higher-resolution (e.g., ALMA) data are needed to study the small-scale structure of this massive clump.

\section{APPENDIX\\
Channel maps of $^{13}$CO (1-0) line emission}
\renewcommand{\thefigure}{A\arabic{figure}}

\setcounter{figure}{0}

\begin{figure*}[tbh!]
\centering
\includegraphics[angle=-90,scale=0.7]{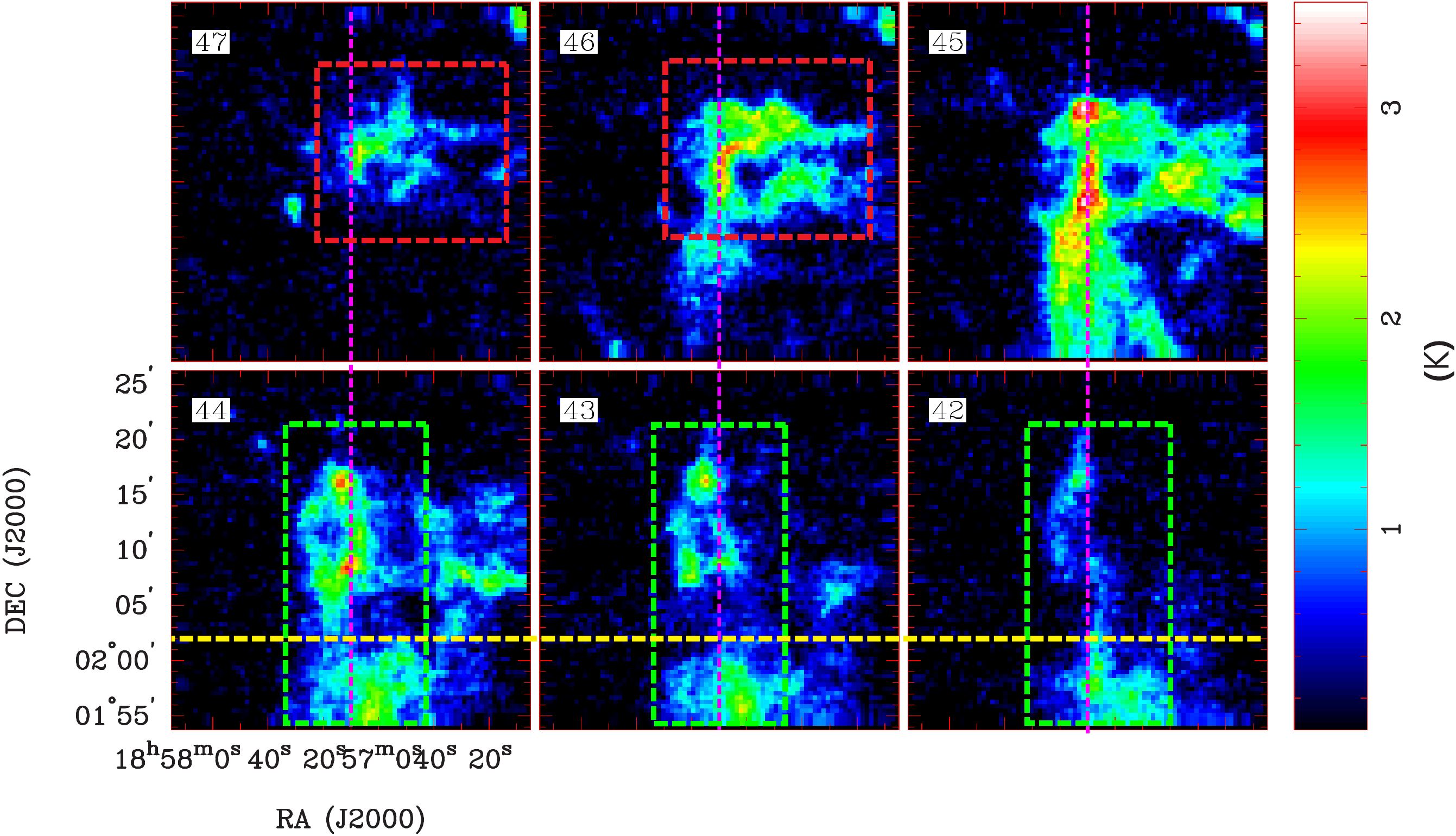}
\caption{Channel maps of the $^{13}$CO (1-0) line emission. The magenta vertical lines represent the long axis of the main filament. The red dashed boxes mark the cloud gas with redshifted velocity, while the green dashed boxes mark the cloud gas with blueshifted velocity. The yellow dashed line divides G035.39-main into two parts (southern and northern parts). The velocities of each panel are shown in the upper-left corners. \label{chan} }
\end{figure*}

Figure \ref{chan} presents the channel maps of $^{13}$CO (1-0) line emission for the $\sim$45 km~s$^{-1}$ component. From the channel maps, we identify two velocity-coherent clouds whose spatial distributions are distinctly different. The western cloud (hereafter we denote it as G035.39-west) with redshifted velocity is mainly distributed in the north-west part of the images, as marked by the red dashed boxes in the 46 km~s$^{-1}$ and 47 km~s$^{-1}$ channel maps. The cloud with blueshifted velocity is a long ($\sim$20 pc) filamentary cloud (hereafter denoted as G035.39-main) distributed along the north to south direction, as marked by the green dashed boxes in the 42 km~s$^{-1}$, 43 km~s$^{-1}$ and 44 km~s$^{-1}$ channel maps. The G035.39-main cloud is divided into two parts by the yellow dashed line in the channel maps. The two parts are connected in velocity space. The high velocity emissions of G035.39-west and G035.39-main clouds are well separated by the magenta dashed line in the channel maps which marks the major axis of the massive filament \textit{$F_{M}$}. The line emission of G035.39-west at its high velocity (47 km~s$^{-1}$) channel is mainly distributed to the west of \textit{$F_{M}$}. On the other hand, the high velocity (43-44 km~s$^{-1}$) emission of the northern part of G035.39-main is mainly distributed to the east of the \textit{$F_{M}$}. The brightest $^{13}$CO (1-0) line emission is in $\sim$45 km~s$^{-1}$ channel, where two clouds overlap.

\section*{Acknowledgment}
\begin{acknowledgements}

We thank the referee, Andrea Bracco, for very valuable comments and suggestions which have improved the content and clarity of this paper. Tie Liu is supported by EACOA fellowship. P.S. Li is supported by NASA ATP grant NNX13AB84G. MJ acknowledges the support of the Academy of Finland Grant No. 285769. JMa acknowledges the support of ERC-2015-STG No. 679852 RADFEEDBACK. C.-P. Zhang is supported by the National Natural Science Foundation of China 11703040. J. Yuan is supported by the National Natural Science Foundation of China through grant 11503035. K.Q. acknowledges the support from National Natural Science Foundation of China (NSFC) through grants NSFC 11473011 and NSFC 11590781. Ke Wang is supported by grant WA3628-1/1 of the German Research Foundation (DFG) through the priority program 1573 (``Physics of the Interstellar Medium''). CWL was supported by Basic Science
Research Program though the National Research Foundation
of Korea (NRF) funded by the Ministry of Education, Science, and
Technology (NRF-2016R1A2B4012593). SPL and KMP acknowledge support from the Ministry of Science and Technology of Taiwan with Grant MOST 106-2119-M-007-021-MY3. This research was partly supported by the OTKA grant NN-111016. This work was carried out in part at the Jet Propulsion Laboratory, which is operated for NASA by the California Institute of Technology. W.K. was supported by the Basic Science Research Program through the National Research Foundation of Korea (NRF-2016R1C1B2013642). The James Clerk Maxwell Telescope is operated by the East Asian Observatory on behalf of The National Astronomical Observatory of Japan; Academia Sinica Institute of Astronomy and Astrophysics; the Korea Astronomy and Space Science Institute; the Operation, Maintenance and Upgrading Fund for Astronomical Telescopes and Facility Instruments, budgeted from the Ministry of Finance (MOF) of China and administrated by the Chinese Academy of Sciences (CAS), as well as the National Key R\&D Program of China (No. 2017YFA0402700). Additional funding support is provided by the Science and Technology Facilities Council of the United Kingdom and participating universities in the United Kingdom and Canada. The KVN is a facility operated by the Korea Astronomy and Space Science Institute.

\software{Starlink software \citep{curr14,jenn13,chap13}}

\end{acknowledgements}


\clearpage

\begin{thebibliography}

\bibitem[Alina et al.(2017)]{alina17}Alina, D., Ristorcelli, I., Montier, L., et al., 2017, eprint arXiv:1712.09325

\bibitem[Andr\'{e} et al.(2014)]{and14}Andr\'{e}, P., Di Francesco, J., Ward-Thompson, D., et al. 2014, Protostars and Planets VI, Henrik Beuther, Ralf S. Klessen, Cornelis P. Dullemond, and Thomas Henning (eds.), University of Arizona Press, Tucson, 914 pp., p.27-51

\bibitem[Auddy, Basu, \& Kudoh(2016)]{Auddy16}Auddy, S., Basu, S., Kudoh, T., et al., 2016, \apj, 831, 46

\bibitem[Bertoldi \& McKee(1992)]{bert92}Bertoldi F., McKee C. F., 1992, \apj, 395, 140

\bibitem[Barnes et al.(2016)]{bar16}Barnes, A. T., Kong, S., Tan, J. C., et al. 2016, \mnras, 458, 1990

\bibitem[Beuther et al.(2013)]{beut13} Beuther, H., Linz, H., Tackenberg, J., et al.\ 2013, \aap, 553, A115

\bibitem[Berry(2007)]{berry07}Berry D. S., Reinhold K., Jenness T., Economou F., 2007, in Shaw R. A.,
Hill F., Bell D. J., eds, ASP Conf. Ser. Vol. 376, Astronomical Data
Analysis Software and Systems XVI. Astron. Soc. Pac., San Francisco,
p. 425

\bibitem[Berry(2015)]{berry15}Berry D. S., 2015, Astron. Comput., 10, 22

\bibitem[Bisbas et al.(2018)]{bisbas18}Bisbas, T. G., Tan, J. C., Csengeri, T., et al., 2018, \mnras, in press, arXiv:1803.00566

\bibitem[Chandrasekhar \& Fermi(1953)]{chandr53}Chandrasekhar, S., \& Fermi, E. 1953, \apj, 118, 116

\bibitem[Chapman et al.(2011)]{chap11}Chapman, N. L., Goldsmith, P. F., Pineda, J. L., et al. 2011, \apj, 741, 21

\bibitem[Chapin et al.(2013)]{chap13}Chapin, E. L., Berry, D. S., Gibb, A. G., et al. 2013, \mnras, 430, 2545

\bibitem[Contreras et al.(2016)]{cont16}Contreras, Y., Garay, G., Rathborne, J. M., et al. 2016, \mnras, 456, 2041

\bibitem[Contreras et al.(2017)]{cont17} Contreras, Y., Rathborne, J.~M., Guzman, A., et al.\ 2017, \mnras, 466, 340

\bibitem[Contreras et al.(2018)]{cont18}Contreras, Y., Sanhueza, P.,Jackson, James M., et al., 2018, \apj, in press, eprint arXiv:1805.01802

\bibitem[Cox et al.(2016)]{cox16}Cox, N. L. J., Arzoumanian, D., Andr\'{e}, Ph., et al. 2016, \aap, 590, 110

\bibitem[Crutcher et al.(2004)]{crut04}Crutcher, R. M., Nutter, D. J., Ward-Thompson, D., \& Kirk, J. M. 2004, \apj, 600, 279

\bibitem[Crutcher et al.(2010)]{crut10}Crutcher, R. M., Wandelt, B., Heiles, C., et al. 2010, \apj, 725, 466

\bibitem[Currie et al.(2014)]{curr14}Currie, M. J., Berry, D. S., Jenness, T., et al. 2014, in Astronomical Society of the Pacific Conference Series, Vol. 485, Astronomical Data Analysis Software and Systems XXIII, ed. N. Manset \& P. Forshay, 391

\bibitem[Davis \& Greenstein(1951)]{davis51}Davis, L. J. \& Greenstein, J. L., 1951, \apj, 114, 206

\bibitem[Eden et al.(2017)]{Eden17}Eden, D. J., Moore, T. J. T., Plume, R., et al. 2017, \mnras, 469, 2163

\bibitem[Federrath(2016)]{fed16}Federrath C., 2016, \mnras, 457, 375

\bibitem[Fiege \& Pudritz(2000)]{Fiege00}Fiege, J. D., \& Pudritz, R. E. 2000, \mnras, 311, 85

\bibitem[Friberg et al.(2016)]{frib16}Friberg, P., Bastien, P., Berry, D., et al. 2016, in
Proc. SPIE, Vol. 9914, Millimeter, Submillimeter, and
Far-Infrared Detectors and Instrumentation for
Astronomy VIII, 991403

\bibitem[Fuller et al.(2005)]{full05}Fuller, G. A., Williams, S. J., \& Sridharan, T. K., 2005, \aap, 442, 949

\bibitem[Giannetti et al.(2014)]{Gian14}Giannetti, A., Wyrowski, F., Brand, J., et al., 2014, \aap, 570, 65

\bibitem[Girart et al.(2013)]{gira13}Girart, J. M., Frau, P., Zhang, Q., et al., 2013, \apj, 772, 69

\bibitem[Gomez, Vazquez-Semadeni, \& Zamora-Aviles(2018)]{Gomez18}Gomez, G. C., Vazquez-Semadeni, E., Zamora-Aviles, M., 2018, submitted to \mnras, eprint arXiv:1801.03169

\bibitem[Guzm{\'a}n et al.(2015)]{Guzm15} Guzm{\'a}n, A.~E., Sanhueza, P., Contreras, Y., et al.\ 2015, \apj, 815, 130

\bibitem[Jenness et al.(2013)]{jenn13}Jenness, T., Chapin, E. L., Berry, D. S., et al. 2013, SMURF: SubMillimeter User Reduction Facility, Astrophysics Source Code Library., , , ascl:1310.007:

\bibitem[Jin et al.(2016)]{jin16}Jin, M., Lee, J.-E., Kim, K.-T., et al., 2016, \apjs, 225, 21

\bibitem[Jim\'{e}nez-Serra et al.(2010)]{jim10}Jim\'{e}nez-Serra, I., Caselli, P., Tan, J. C., et al., 2010, \mnras, 406, 187

\bibitem[Jim\'{e}nez-Serra et al.(2014)]{jim14}Jim\'{e}nez-Serra, I., Caselli, P., Fontani, F., et al.  2014, \mnras, 439, 1996

\bibitem[Juvela et al.(2010)]{juvela10}Juvela, M., Ristorcelli, I., Montier, L. A., et al., 2010, \aap, 518, L93

\bibitem[Juvela et al.(2012)]{juvela12}Juvela, M., Ristorcelli, I., Pagani, L., et al. 2012, \aap, 541, 12

\bibitem[Juvela et al.(2018a)]{juvela18a}Juvela, M., He, J., Pattle, K., et al. 2018a, \aap, 612, 71

\bibitem[Hacar et al.(2013)]{hacar13}Hacar, A., Tafalla, M., Kauffmann, J., Kov\'{a}cs, A., 2013, \aap, 554, 55

\bibitem[Hacar et al.(2016)]{hacar16}Hacar, A., Kainulainen, J., Tafalla, M., Beuther, H., Alves, J., 2016, \aap, 587, 97

\bibitem[Henshaw et al.(2014)]{henshaw14}Henshaw, J. D., Caselli, P., Fontani, F., Jim\'{e}nez-Serra, I., \& Tan, J. C. 2014, \mnras, 440, 2860

\bibitem[Henshaw et al.(2013)]{henshaw13}Henshaw, J. D., Caselli, P., Fontani, F., et al. 2013, \mnras, 428, 3425

\bibitem[Henshaw et al.(2016)]{henshaw16}Henshaw, J. D., Caselli, P., Fontani, F., et al. 2016, \mnras, 463, 146

\bibitem[Henshaw et al.(2017)]{henshaw17}Henshaw, J. D., Jim\'{e}nez-Serra, I., Longmore, S. N., et al. 2017, \mnras, 464, 31

\bibitem[Heitsch et al.(2001)]{heit01}Heitsch, F., Zweibel, E. G., Mac Low, M.-M., Li, P., \& Norman, M. L. 2001, \apj, 561, 800

\bibitem[Hernandez \& Tan(2015)]{hern15}Hernandez, A. K. \& Tan, J. C., 2015, \apj, 809, 154	

\bibitem[Holland et al.(2013)]{holland13}Holland, W. S., Bintley, D., Chapin, E. L., et al. 2013, \mnras, 430, 2513

\bibitem[Inoue \& Fukui(2013)]{Inoue13}Inoue, T., \& Fukui, Y. 2013, \apj, 774L, L31

\bibitem[Jackson et al.(2010)]{jack10}Jackson, J. M., Finn, S. C., Chambers, E. T., et al. 2010, \apj, 791L, 185

\bibitem[Kainulainen \& Tan(2013)]{kai13}Kainulainen, J. \& Tan, J. C., 2013, \aap, 549, 53

\bibitem[Kauffmann \& Pillai(2010)]{kauf10}Kauffmann, J., \& Pillai, T. 2010, \apjl, 723, L7

\bibitem[Kim et al.(2011)]{kim11}Kim, K.-T., Byun, D.-Y., Je, D.-H., et al., 2011, JKAS, 44, 81

\bibitem[Kim et al.(2017)]{kim17}Kim, J., Lee, J.-E., Liu, T., et al., 2017, \apjs, 231, 9

\bibitem[Kwon et al.(2018)]{kwon18}Kwon, J., Doi, Y., Tamura, M., et al., 2018, \apj, 859, 4

\bibitem[Kirk et al.(2013)]{kirk13}Kirk, H., Myers, P. C., Bourke, T. L., et al. 2013, \apj, 766, 115

\bibitem[Klassen et al.(2017)]{klass17}Klassen, M., Pudritz, R. E., Kirk, H., 2017, \mnras, 465, 2254

\bibitem[Koch, Tang \& Ho(2012)]{koch12}Koch, P. M., Tang, Y.-W. \& Ho, P. T. P., 2012, \apj, 747, 79

\bibitem[Koch et al.(2014)]{koch14}Koch, P. M., Tang, Y.-W., Ho, P. T. P., et al. 2014, \apj, 797, 99

\bibitem[Koch \& Rosolowsky(2015)]{koch15}Koch, E. W., \& Rosolowsky, E. W. 2015, \mnras, 452, 3435

\bibitem[Li et al.(2009)]{li09}Li, H.-B., Dowell, C. D., Goodman, A., Hildebrand, R., Novak, G., 2009, \apj, 704, 891

\bibitem[Li, McKee \& Klein(2015)]{lips15}Li, P. S., McKee, C. F. \& Klein, R. I., 2015, \mnras, 452, 2500

\bibitem[Li et al.(2015)]{lihb15}Li, H.-B., Yuen, K. H., Otto, F., et al., 2015, \nat, 520, 518

\bibitem[Liu et al.(2011a)]{liu11a}Liu, T., Wu, Y., Zhang, Q., et al., 2011a, \apj, 728, 91

\bibitem[Liu et al.(2011b)]{liu11b}Liu, T., Wu, Y., Liu, S.-Y., et al., 2011b, \apj, 730, 102

\bibitem[Liu, Wu \& Zhang(2012)]{liu12}Liu, T., Wu, Y., \& Zhang, H., 2012, \apjs, 202, 4

\bibitem[Liu Wu \& Zhang(2013a)]{liu13a}Liu, T., Wu, Y., \& Zhang, H., 2013a, \apj, 776, 29

\bibitem[Liu et al.(2013b)]{liu13b}Liu, T., Wu, Y., Wu, J., et al., 2013b, \mnras, 436, 1335

\bibitem[Liu, Wu \& Zhang(2013c)]{liu13c}Liu, T., Wu, Y., \& Zhang, H., 2013c, \apj, 775L, 2L

\bibitem[Liu et al.(2015)]{liu15}Liu, T., Wu, Y., Mardones, D., et al., 2015, PKAS, 30, 79

\bibitem[Liu et al.(2016a)]{liu16a}Liu, T., Zhang, Q., Kim, K.-T., et al. 2016a, \apj, 824, 31

\bibitem[Liu et al.(2016b)]{liu16b}Liu, T., Kim, K.-T., Yoo, H., et al., 2016b, \apj, 829, 59

\bibitem[Liu et al.(2016c)]{liu16c}Liu, T., Zhang, Q., Kim, K. -T., et al. 2016c, \apjs, 222, 7

\bibitem[Liu et al.(2017)]{liu17}Liu, T., Lacy, J., Li, P. S., et al. 2017, \apj, 849, 25

\bibitem[Liu et al.(2018)]{liu18}Liu, T., Kim, K.-T., Juvela, M., et al. 2018, \apjs, 234, 28

\bibitem[Li, Klein \& McKee(2017)]{lips17}Li, P. S., Klein, R. I., \& McKee, C. F., 2017, \mnras, 473, 4220

\bibitem[Lu et al.(2018)]{lu18}Lu, X., Zhang, Q., Liu, H. B., et al. 2018, \apj, 855, 9

\bibitem[Malinen et al.(2016)]{mali16}Malinen, J., Montier, L., Montillaud, J., et al. 2016, \mnras, 460, 1934

\bibitem[Meng, Wu \& Liu(2013)]{meng13}Meng, F., Wu, Y., Liu, T., et al. 2013, \apjs, 209, 37

\bibitem[Molinari et al.(2010)]{Molinari2010}Molinari, S., Swinyard, B., Bally, J., et al., 2010, \aap, 518, L100

\bibitem[Montillaud et al.(2015)]{mont15}Montillaud, J., Juvela, M., Rivera-Ingraham, A., 2015, \aap, 584, 92

\bibitem[Moore et al.(2015)]{Moore15}Moore, T. J. T., Plume, R., Thompson, M. A., et al., 2015, \mnras, 453, 4264

\bibitem[Nguyen Luong et al.(2011)]{Nguyen11}Nguyen Luong, Q., Motte, F., Hennemann, M., et al., 2011, \aap, 535, 76

\bibitem[Ostriker(1964)]{ostri64}Ostriker, J. 1964, \apj, 140, 1056

\bibitem[Ostriker, Stone, \& Gammie(2001)]{ostr01}Ostriker, E. C., Stone, J. M., \& Gammie, C. F. 2001, \apj, 546, 980

\bibitem[Palmeirim et al.(2013)]{Palmeirim2013}Palmeirim, P., Andr\'{e}, Ph., Kirk, J., et al. 2013, \aap, 550, 38

\bibitem[Panopoulou et al.(2017)]{pan17}Panopoulou, G. V., Psaradaki, I., Skalidis, R., et al. 2017, \mnras, 466, 2529

\bibitem[Pattle et al.(2017)]{pattle17}Pattle, K., Ward-Thompson, D., Berry, D., et al., 2017, \apj, 846, 122

\bibitem[Pillai et al.(2011)]{pillai11}Pillai T., Kauffmann J., Wyrowski F., Hatchell J., Gibb A. G., Thompson M. A., 2011, \aap, 530, A118

\bibitem[Pillai et al.(2015)]{pillai15}Pillai T., Kauffmann J., Tan J. C., Goldsmith P. F., Carey S. J., Menten K. M., 2015, \apj, 799, 74

\bibitem[Planck Collaboration XXIII(2011)]{planck11a}Planck Collaboration, Ade, P. A. R., Aghanim, N., et al. 2011, \aap, 536, 23

\bibitem[Planck Collaboration XXII(2011)]{planck11b}Planck Collaboration, Ade, P. A. R., Aghanim, N., et al. 2011, \aap, 536, 22

\bibitem[Planck Collaboration XXXV(2016)]{planck16a}Planck Collaboration, Ade, P. A. R., Aghanim, N.,  et al. 2016, \aap, 586, 138

\bibitem[Planck Collaboration XXXII(2016)]{planck16b}Planck Collaboration, Adam, R., Ade, P. A. R., et al., 2016, \aap, 586, 135

\bibitem[Planck Collaboration XXVIII(2016)]{planck16c}Planck Collaboration, Ade, P. A. R., Aghanim, N., et al. 2016, \aap, 594, 28

\bibitem[Peretto et al.(2013)]{pere13}Peretto, N., Fuller, G. A., Duarte-Cabral, A., et al. 2013, \aap, 555, 112

\bibitem[Qin et al.(2016)]{qin16}Qin, S.-L., Schilke, P., Wu, J., et al., 2016, \mnras, 456, 2681

\bibitem[Qiu et al.(2013)]{qiu13}Qiu, K., Zhang, Q., Menten, K. M., et al., 2013, \apj, 779, 182

\bibitem[Qiu et al.(2014)]{qiu14}Qiu, K., Zhang, Q., Menten, K. M., et al., 2014, \apj, 794L, 18

\bibitem[Ren et al.(2012)]{ren12}Ren, Z., Wu, Y., Zhu, M., et al. 2012, \mnras, 422, 1098

\bibitem[Rivera-Ingraham et al.(2016)]{rivera16}Rivera-Ingraham, A., Ristorcelli, I., Juvela, M., et al., 2016, \apj, 591, 90

\bibitem[Rivera-Ingraham et al.(2017)]{rivera17}Rivera-Ingraham, A., Ristorcelli, I., Juvela, M., et al., 2017, \apj, 601, 94

\bibitem[Santos et al.(2016)]{sant16}Santos, F. P., Busquet, G., Franco, G. A. P., et al., 2016, \apj, 832, 186

\bibitem[Sanhueza et al.(2010)]{sanh10}Sanhueza, P., Garay, G., Bronfman, L., et al. 2010, \apj, 715, 18

\bibitem[Sanhueza et al.(2012)]{sanh12}Sanhueza, P., Jackson, J. M., Foster, J. B., et al., 2012, \apj, 756, 60

\bibitem[Sanhueza et al.(2013)]{sanh13} Sanhueza, P., Jackson, J.~M., Foster, J.~B., et al.\ 2013, \apj, 773, 123

\bibitem[Sanhueza et al.(2017)]{sanh17}Sanhueza, P., Jackson, J. M., Zhang, Q., et al., 2017, \apj, 841, 97

\bibitem[Shirley(2015)]{shir15}Shirley, Y. L., 2015, \pasp, 127, 299

\bibitem[Simon et al.(2006)]{simon06}Simon R., Rathborne J. M., Shah R. Y., Jackson J. M., Chambers E. T., 2006, \apj, 653, 1325

\bibitem[Soam et al.(2018)]{soam18}Soam, A., Pattle, K., Ward-Thompson, D., et al., ApJ in press, arXiv:1805.06131

\bibitem[Sokolov et al.(2017)]{soko17}Sokolov, V., Wang, K., Pineda, J. E., et al. 2017, \aap, 606, 133

\bibitem[Soler et al.(2017)]{sole17a}Soler, J. D., Ade, P. A. R., Angil\`{e}, F. E., et al., 2017, \aap, 603, 64

\bibitem[Soler \& Hennebelle(2017)]{sole17}Soler, J. D. \& Hennebelle, P., 2017, \aap, 607, 2

\bibitem[Takahira, Tasker \& Habe(2014)]{taka14}Takahira, K., Tasker, E. J., \& Habe, A. 2014, \apj, 792, 63

\bibitem[Tan et al.(2013)]{Tan13} Tan, J.~C., Kong, S., Butler, M.~J., Caselli, P., \& Fontani, F.\ 2013, \apj, 779, 96

\bibitem[Tang et al.(2018a)]{tang18a}Tang, Y.-W., Koch, P.M., Peretto, N., Novak, G., Duarte-Cabral, A. and Chapman, N.L. 2018a, \apj, submitted

\bibitem[Tang et al.(2018b)]{tang18b}Tang, M., Liu, T., Qin, S.-L., et al. 2018b, \apj, 856, 141

\bibitem[Tatematsu et al.(2017)]{tat17}Tatematsu, K., Liu, T., Ohashi, S., et al. 2017, \apjs, 228, 12

\bibitem[Traficante et al.(2015)]{traf15}Traficante, A., Fuller, G.~A., Peretto, N., Pineda, J.~E., \& Molinari, S.\ 2015, \mnras, 451, 3089

\bibitem[Van der Tak et al.(2007)]{van07}Van der Tak, F.F.S., Black, J.H., Sch\"{o}ier, F.L., Jansen, D.J., van Dishoeck, E.F. 2007, \aap, 468, 627

\bibitem[Wang et al.(2016)]{wang16}Wang, K., Testi, L., Burkert, A., et al. 2016, \apjs, 226, 9

\bibitem[Ward-Thompson et al.(2017)]{ward17}Ward-Thompson, D., Pattle, K., Bastien, P., et al. 2017, \apj, 842, 66

\bibitem[Wu \& Evans(2003)]{wu03}Wu, J. \& Evans, N. J., II, 2003, \apj, 592L, 79

\bibitem[Wu et al.(2007)]{wu07}Wu, Y., Henkel, C., Xue, R.,et al., 2007, \apj, 669L, 37

\bibitem[Wu et al.(2009)]{wu09}Wu, Y., Qin, S.-L., Guan, X., et al. 2009, \apj, 697L, 116

\bibitem[Wu et al.(2012)]{wu12}Wu, Y., Liu, T., Meng, F., et al. 2012, \apj, 756, 76

\bibitem[Wu et al.(2014)]{wu14}Wu, Y., Liu, T., Qin, S.-L., 2014, \apj, 791, 123

\bibitem[Wu et al.(2015)]{wu15}Wu, B., Van Loo, S., Tan, J. C., et al. 2015, \apj, 811, 56

\bibitem[Wu et al.(2017)]{wu17}Wu, B., Tan, J. C., Nakamura, F., et al. 2017, \apj, 835, 137

\bibitem[Yi et al.(2018)]{Yi18}Yi, H.-W., Lee, J.-E., Liu, T., et al., 2018, ApJS in press, arXiv:1805.05738

\bibitem[Yuan et al.(2016)]{yuan16}Yuan, J., Wu, Y., Liu, T., et al., 2016, \apj, 820, 37

\bibitem[Yuan et al.(2017)]{yuan17}Yuan, J., Wu, Y., Ellingsen, S. P., et al. 2017, \apjs, 231, 11

\bibitem[Yuan et al.(2018)]{yuan18}Yuan, J., Li, J.-Z., Wu, Y., et al. 2018, \apj, 852, 12

\bibitem[Zhang et al.(2014)]{zhang14}Zhang, Q., Qiu, K., Girart, J. M., et al. 2014, \apj, 792, 116

\bibitem[Zhang et al.(2017)]{zhang17}Zhang, C.-P., Yuan, J.-H., Li, G.-X., et al., 2017, \aap, 598, 76

\bibitem[Zhang et al.(2016)]{zhang16}Zhang, T., Wu, Y., Liu, T., et al. 2016, \apjs, 224,43

\bibitem[Zhang et al.(2018)]{zhang18}Zhang, C.-P., Liu, T., Yuan, J., et al. 2018, ApJS, in press, arXiv:1805.03883

\bibitem[Zhou et al.(1993)]{zhou93}Zhou, S., Evans, N. J., II; Koempe, C., et al., 1993, \apj, 404, 232

\end{thebibliography}
\end{document}